\documentclass[
        english,
        reqno,            
        nonumbib,         
        twocolid,         
]{nrc2}

\usepackage{natbib}
\usepackage{verbatim}

\usepackage{graphicx}       

\usepackage{amsmath}        
\usepackage{amssymb}        

\usepackage{url}            
\NRCurl{url}

\def\beq{\begin{eqnarray}}
\def\eeq{\end{eqnarray}}

\def\beqn{\begin{eqnarray*}}  
\def\eeqn{\end{eqnarray*}}

\def\E{{\rm E}}
\def\Var{{\rm Var}}

\def\Pr{\rm Pr}

\def\bulk{{\rm bulk}}
\def\ind{{\rm ind}}
\def\HSI{{\rm HSI}}
\def\year{{\rm year}}
\def\AR{{\rm AR}}
\def\cc{{\rm cc}}

\def\hatt{\widehat}

\def\eps{\varepsilon}
\def\half{\hbox{$1\over2$}}

\def\rootn{\sqrt{n}}

\def\true{{\rm true}}

\def\AIC{{\rm AIC}}
\def\BIC{{\rm BIC}}

\volyear{XX}{2015}               
\journal{Can. J. Fish. Aquat. Sci.}                       
\journalcode{cjfas}                   
\filenumber{}                    

\received{}                      
\revreceived{}                   
\accepted{}                      
\revaccepted{}                   

\begin{document}



\title{Recent advances in statistical methodology applied to
the Hjort liver index time series (1859-2012) and associated
influential factors}


\author[G. H. Hermansen]{Gudmund H. Hermansen} 
\address[UiO1]{Department of Mathematics, University of Oslo}  
\correspond{gudmund.hermansen@gmail.com}

\author[N. L. Hjort]{Nils Lid Hjort}
\address[UiO1]

\author[O. S. Kjesbu]{Olav S. Kjesbu}
\address{Institute of Marine Research (IMR) 
   and Hjort Centre for Marine Ecosystem Dynamics, Bergen,  
   and Centre for Ecological and Evolutionary Synthesis (CEES), 
   Department of Biosciences, University of Oslo}

\shortauthor{Hermansen, Hjort, and Kjesbu} 

\begin{abstract}
Certain recent advances in statistical methodology 
have promising potential for fruitful use in general biology 
and the fisheries sciences. This paper reviews and discusses 
some of the relevant themes, including accurate modelling 
via focused model selection techniques, dynamic goodness-of-fit
testing of processes evolving over time, finding break points
for phenomena experiencing changes, prediction uncertainty,
and optimal combination of information across diverse sources
via confidence distributions.  
The methods are illustrated for the Hjort liver quality index
time series. 
Its roots lie in the classic \cite{JHjort14}, where liver quality 
of the Atlantic cod {\it (Gadus morhua)} for 1880--1912 
is reported on and studied,
along with related factors, making it one of the first 
teleost time series ever published. 
Diligent work by \cite{Kjesbuetal14A}, involving both 
archival and calibration efforts, have extended the series
both backwards and forwards in time, to 1859--2012, 
yielding one of the longest time series of marine science. 
Our study offers a detailed examination of this series and 
how it relates to and interacts with associated factors,
including Kola winter temperatures, length distribution
parameters, cod mortality, and a certain index related 
to availability of food.  

\keywords{
Atlantic cod, 
focused information criteria,
Johan Hjort, 
liver quality index, 
model selection,
prediction,
time series modelling}
\end{abstract}

\begin{resume}
\vspace{4.00cm}

\traduit 
\end{resume}

\maketitle			

\section*{Introduction}
\label{section:intro}

The first four chapters of the classic \cite{JHjort14} are 
essentially occupied with the {\it quantity} of fish (specifically, 
the herring and the cod), the associated underlying causes 
driving its fluctuations, etc. He was however also 
concerned with what he terms the {\it quality} of fish, 
and devotes most of the book's Chapter 5 to discuss
how this can reasonably be defined and measured, 
also attempting to identify factors involved as it varies
from year to year. He proposed using the liver
quality index ``no.~of hectolitres of liver pr.~1,000 skrei''
for such a purpose, and established a time series 
of such measurements for the northeast Arctic cod 
(skrei, {\it Gadus morhua}), for the years 1880--1912. 
This is arguably one of the first comprehensive teleost time series
ever published. A few points of the same series were used 
and studied in \citet[Ch.~7.3]{HellandHansenNansen1909}. 

\begin{figure*}[ht]
\centering
\includegraphics[width=0.90\textwidth]{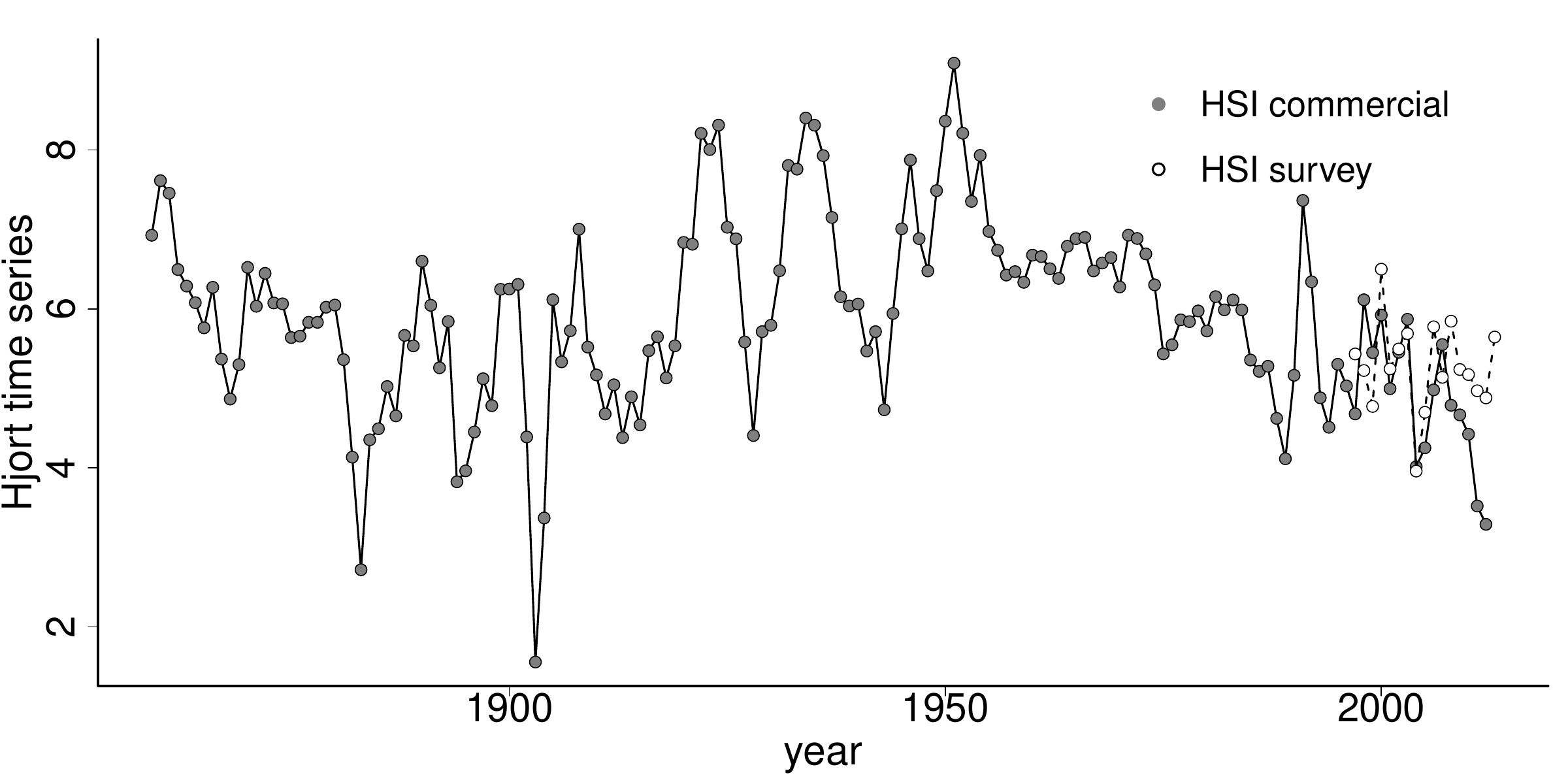}
\caption{Hjort time series from commercial fishing and 
IMR's Lofoten surveys (the latter from 1997 onwards), see 
\cite{Kjesbuetal14A}.}
\label{figure:Hjort}
\end{figure*}

Work summarised in \cite{Kjesbuetal14A} has made it possible 
to extend this liver quality time series both backwards and 
forwards in time, using data from both Fisheries Statistics
(1859--1990) and R{\aa}fiskelaget (1991--2012) along with
further archival efforts. 
In this process it has been found fruitful to pass from the 
somewhat crude volumetric based hecto-litres of liver 
per 1,000 fish used earlier to a more naturally standardised 
measurement called the hepatosomatic index HSI. In bulk form
this HSI is defined as total amount of liver (in kg) 
divided by total amount of fish (in kg). 
Passing from volumetric to weight scale has been
achieved via an essentially linear relationship
learned from regression analysis; see 
\citet[eq.~(2)]{Kjesbuetal14A}. 
The resulting HSI data, perhaps the longest-running marine science 
time series there is, are displayed in Figure \ref{figure:Hjort}. 
These authors also investigate the
extent to which the Kola temperature, where annual measurements
are available from 1900 and with more detailed monthly average
temperatures from 1921 onwards, see \cite{Boitsovetal12}, 
can be seen to influence or interact with the HSI. 
Also other covariates and their degree of relevance 
for the HSI are reviewed and examined in \citet{Kjesbuetal14B}, 
including fat content and body size.  

That the HSI carries important biological information
for the quality of a stock of fish has been demonstrated
in several publications, from \cite{HellandHansenNansen1909}
and \cite{JHjort14} onwards. 
The background for this interest in amount of the liver from 
earlier days and up to today rests with the fact that this 
organ not only accumulates fat for subsequent metabolic costs 
in capital breeders (cf.~stored energy) like cod but also 
is a production site for yolk (vitellogenin) and eggshell 
(chorion) material, which are transported by the blood 
to the maturing ovary \citep{TylerSumpter96}. Also, at 
the population level, HSI has shown clear positive links 
to the level of prey, in particular the stock size of 
capelin {\it (Mallotus villosus)}, but also to individual 
fecundity and thereby total egg production \citep{Marshalletal99}.
Further elaborations of the significance of the liver 
in the present context are given in \citet{Sandemanetal08}
and \citet{Kjesbuetal14A}, where it is argued that 
the HSI is the universal expression of investment in liver 
size vs.~body size.  

The aims of this article are two-fold. The first goal 
is to provide more careful statistical analyses of the HSI 
parameter itself, from its definition, interpretation and 
generalisation from bulk index to individual index, 
to aspects of the full 1859--2012 time series. We also
examine degrees of associations with related quantities,
like the Kola temperatures, and provide glimpses into
the future, using data also to predict ahead as opposed
to focusing on understanding the past.  

Our second ambition is to use the opportunity to provide
perspectives on and a brief overview of certain modern and 
relevant developments in statistical methodology. These relate to 
(i) choosing good models for complex phenomena, partly
via the `focused viewpoint' used in the construction of 
certain focused information criteria for model selection, 
see \cite{HjortClaeskens03, ClaeskensHjort03, ClaeskensHjort08, 
HermansenHjort15a, HermansenHjort15b}; 
(ii) assessing adequacy of fit using dynamical monitoring
of log-likelihood maxima and other criteria, useful for 
examining processes evolving over time; 
(iii) the sometimes challenging statistical dividing line 
between `to explain' and `to predict',
cf.~\citet{Breiman01, Shmueli10}; and 
(iv) ways of optimally combining information across
diverse data sources, see \citet{XieSingh13, SchwederHjort15}.

\section*{The HSI index and targets for inference} 
\label{section:theHSIindex}

In \citet[Ch.~5]{JHjort14}, where the liver index is introduced 
as ``no.~of hectolitres of liver pr.~1,000 skrei'', the 
underlying concern and motivation is that of coming to
grips with the {\it quality} of fish, supplementing 
the information of the {\it quantity}. It is clear from 
his discussion that also other and related versions 
of `quality' can be used. In this section we make some
remarks pertaining to the statistical issues involved
when defining and measuring appropriate indexes. 

\subsection*{What is the liver quality index (HSI)?}

The definition mentioned above, used in 
e.g.~\citet[Fig.~107]{JHjort14}, is a practical one,
working in bulk modus, so to speak, without necessitating
detailed examination of each individual fish. It may 
be represented as 
\beq 
\label{eq:HSI2}
\hatt{\HSI}_{\bulk} 
= 100 \times \frac{\textrm{total amount of liver}}
{\textrm{total amount of fish}}
= 100 \times \frac{\bar x}{\bar y},
\eeq 
where $(x_i,y_i)$ represent the weight of liver 
and the total weight for fish no.~$i$ 
and $\bar x$ and $\bar y$ the respective averages,
over a sample of say $n=1000$ fish
(in the Lofoten fishery millions of fish landed 
are actually landed). 
This is the bulk liver index worked with 
in \cite{Kjesbuetal14A}, marked `commercial' 
in Figure \ref{figure:Hjort}. 
An alternative definition, relating more directly to 
the individual fish, is 
\beq
\label{eq:HSI1}
\begin{split}
\hatt{\HSI}_{\ind} 
& =  100 \times \bigg ( \frac{1}{n} \sum_{i = 1}^{n} 
 \frac{\textrm{weight of liver in fish $i$}}
 {\textrm{weight of fish $i$}} \bigg ) \\
& = 100 \times \bigg ( \frac{1}{n} \sum_{i = 1}^{n} \frac{x_{i}}{y_{i}}  \bigg ).
\end{split}
\eeq
This per-fish index has been measured for the years 1997 onwards
as part of IMR's Lofoten research survey, marked `survey' 
in Figure \ref{figure:Hjort}. 

These indexes are both biologically meaningful, 
and are of course related, but not equivalent.
The underlying statistical parameters are respectively 
$\HSI_\bulk=\E\,X/\E\,Y$ for the bulk and $\HSI_\ind=\E\,(X/Y)$
for the per fish index, with $(X,Y)$ denoting liver weight 
and total weight for a randomly selected fish in the population 
in question, and `$\E$' as usual denoting mathematical 
expectation of a random variable.  
The degree to which the two parameters differ 
is determined by aspects of the joint distribution 
of the two quantities $X$ and $Y$, including both
their internal correlation, the spread of the 
distribution of $Y$, and the latter's distance from zero.

\begin{figure}[ht]
\centering
\includegraphics[width=0.45\textwidth]{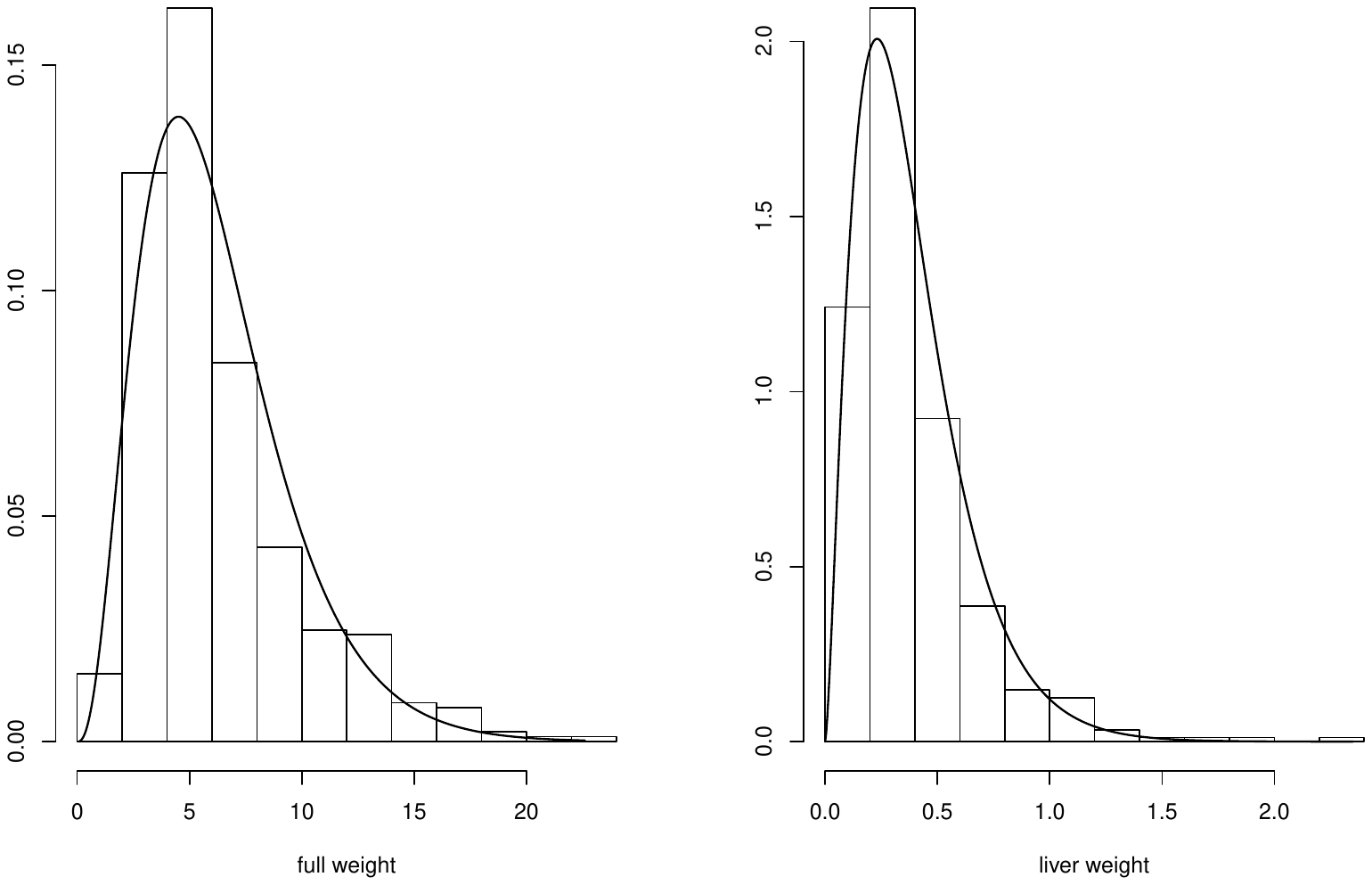}
\caption{Gamma distribution densities fitted to full-weight 
and liver-weight data (both in kg), and with correlation 0.82;
cf.~\cite{Mjangeretal06} for classification of `skrei'
from otolith readings.}
\label{figure:andenes_fittedgammas}
\end{figure}

The time series $\HSI=\HSI_\bulk$ displayed in 
Figure \ref{figure:Hjort} cannot alone provide information 
on values of the individual-based parameter $\HSI_\ind$. 
We may however use research data on individual body metrics 
collated from fishing ports over many years consulting `skrei' 
catches as in \citet{JHjort14} \citep{Kjesbuetal98, Kjesbuetal10}  
to analyse the relevant joint distribution 
of $(X,Y)$, leading also to a mechanism for predicting one 
HSI parameter from the other. Examining these data,
with $n=439$ pairs of $(x_i,y_i)$, 
one learns first that $X$ and $Y$ are separately well
modelled using gamma distributions, and also that 
a five-parameter bivariate gamma model provides 
a fully adequate fit to the joint distribution. 
The model in question takes 
\beq
\label{eq:bivariategamma}
X=G_1^{-1}(\Phi(U),\allowbreak a_1,b_1) 
   {\rm\ and\ }
Y=G_2^{-1}(\Phi(V),a_2,b_2), 
\eeq 
involving the inverse gamma distribution functions 
with parameters $(a_1,b_1)$ and $(a_2,b_2)$ respectively, 
the cumulative standard normal distribution function $\Phi$,
and a standardised binormal pair $(U,V)$ with 
correlation parameter $\rho$. Parameter estimates 
were $2.51,6.52,3.99,\allowbreak 0.63,0.83$, corresponding
in particular to means and standard deviations
6.23 and 3.28 for $x$ and 0.38 and 0.25 for $y$. 
The data fit the estimated gamma densities well 
(Figure \ref{figure:andenes_fittedgammas}). 

\begin{figure}[ht]
\centering
\includegraphics[width=0.45\textwidth]{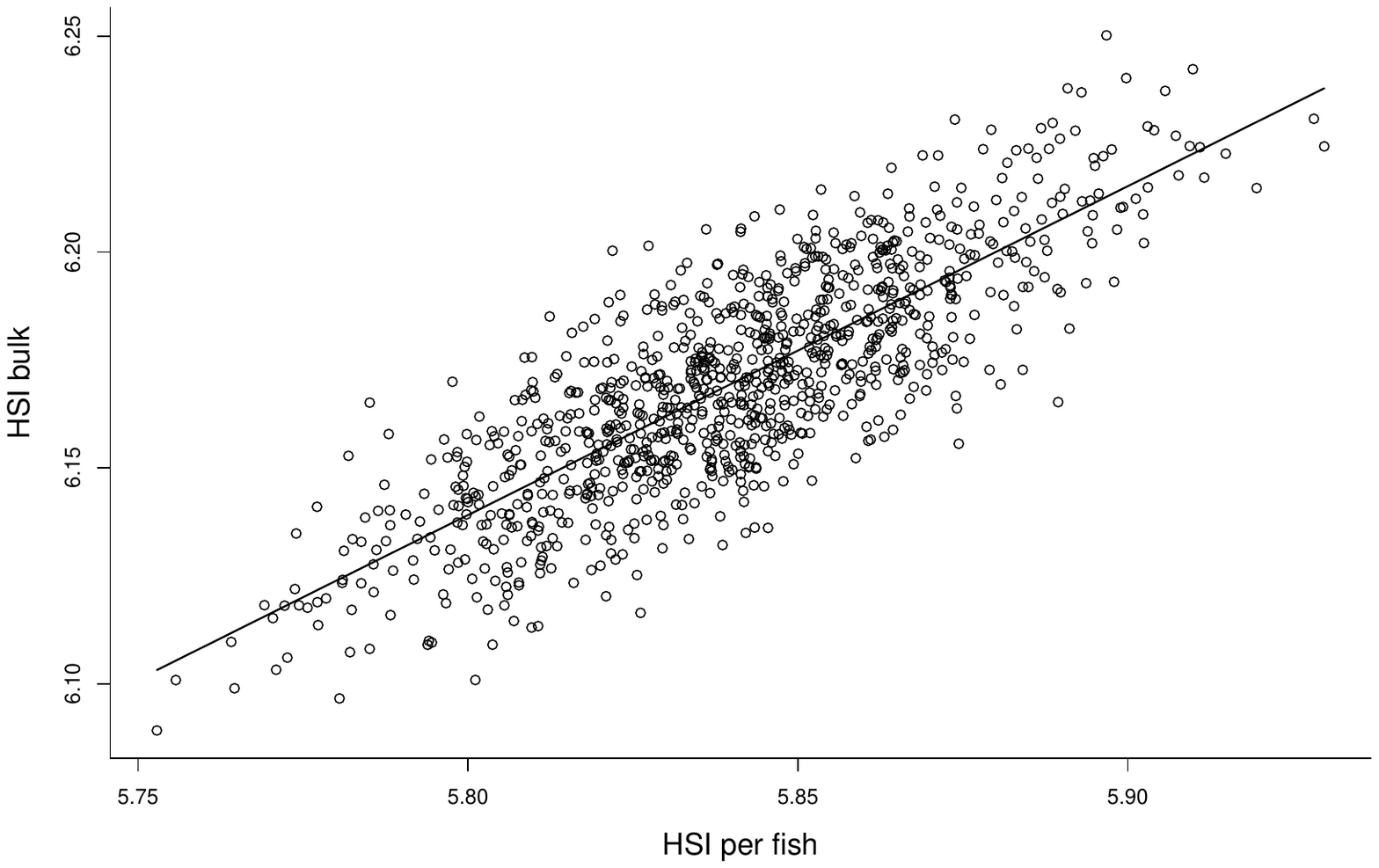}
\caption{Simulated per fish and bulk HSI values using the model 
[\ref{eq:bivariategamma}] 
as per eq.~[\ref{eq:HSI2}] and [\ref{eq:HSI1}]. 
The mean and standard deviations are 5.84 and 0.03 for 
the HSI per fish distribution and 6.17 and 0.03 for the 
HSI bulk distribution and. The correlation is 0.83.}
\label{figure:SimHSI1andHSI2}
\end{figure}

\begin{figure}[ht]
\centering
\includegraphics[width=0.45\textwidth]{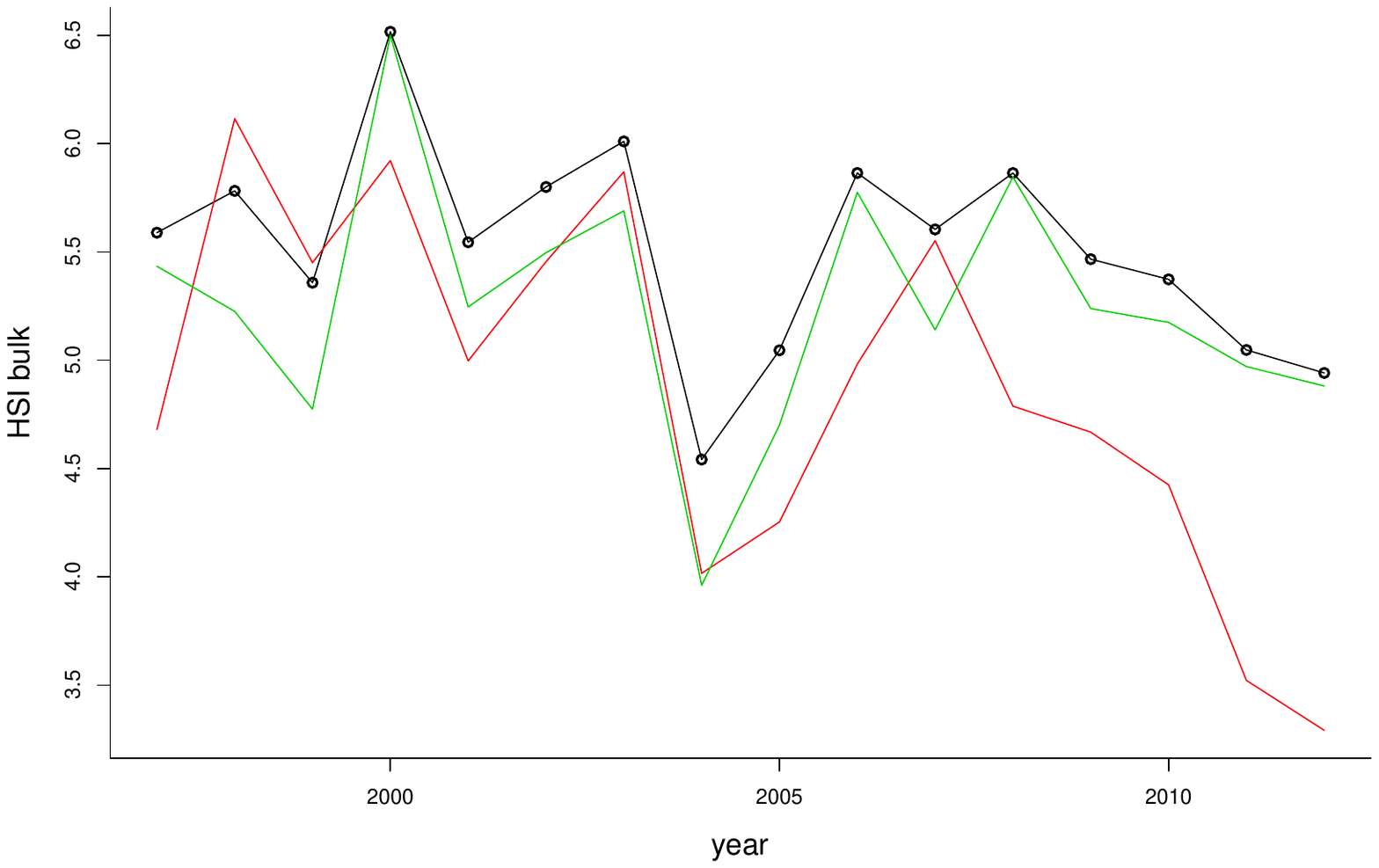}
\caption{Zooming in on 1997 to 2012 of Figure \ref{figure:Hjort}, 
the figure displays both the HSI bulk index from the commercial 
data (red line) and the HSI individual index from the survey data 
(green line), along with the optimal prediction of the HSI bulk index
based on these series (black line), using confidence 
distribution theory.}
\label{figure:commercial_survey_combined}
\end{figure}

We may use the bivariate gamma model [\ref{eq:bivariategamma}]
to infer aspects of the connection between the bulk index 
$\HSI_\bulk$ and the indi\-vidual-fish index $\HSI_\ind$. 
Figure \ref{figure:SimHSI1andHSI2} displays 
simulated pairs from the relevant distribution,
taking for this illustration the sample size $n$
above to be 1000. It leads to a correlation of 0.83 
between the bulk and the per fish HSI indexes. 
This investigation also leads to the tentative formula 
\beq
\label{eq:hsibulkandind}
\HSI_\bulk = 1.581 + 0.786\,\HSI_\ind
\eeq 
for translating the per-fish index to the bulk liver index.
Such a formula would need to be used with care, however,
as its precise coefficients depend on the population
being sampled (as well as, though to a lesser extent, 
on the number of fish in the bulk in question). 
Going back to the Hjort time series displayed in 
Figure \ref{figure:Hjort}, we may zoom in on 1997--2012
to supplement the information there in two ways. 
First, the survey data numbers may be converted 
to bulk index estimates, using [\ref{eq:hsibulkandind}].
Second, the methodology of optimal combination of 
information across data sources, reviewed in a later
section, may be used to provide the best estimates
of the $\HSI_\bulk$ for these years, utilising both
the commercial and the survey data 
(Figure \ref{figure:commercial_survey_combined}).
Note for this application that the resulting estimated
bulk HSI is higher than both the commercial $\HSI_\bulk$ 
numbers and the survey $\HSI_\ind$ numbers, due in part 
to the fact that [\ref{eq:hsibulkandind}] yields
bulk scores higher than individual scores for the 
range of the latter met here (cf.~the scales involved  
in Figure \ref{figure:SimHSI1andHSI2}). 


The relatively high correlation between weight of liver
and that of the fish (it is about 0.83) helps to secure 
(in addition to the relative variances) that the 
difference between $\HSI_\ind$ and $\HSI_\bulk$ are fairly 
close. Simulation studies indicate that there 
is a nearly perfect linear relationship, with 
perfect correspondence in HSI indexes, if the 
correlation is close to one and a approximately 
50\% increase in $\HSI_{\ind}$ compared to $\HSI_{\bulk}$ 
if liver and fish weight are independent. Again, 
this points out the importance of understanding 
the actual target for estimation and also 
the importance of more fundamental analysis, 
which can be used to establish translation formulae 
like the one in [\ref{eq:hsibulkandind}]. 

\subsection*{Modelling HSI as a mixture}

Naturally measuring the per fish liver index for a number 
of specimens takes certain laborious and precision demanding 
efforts. We learn from the detailed analysis reported 
on above, involving the bivariate gamma model [\ref{eq:bivariategamma}], 
that the per fish and bulk liver indexes are strongly related
and that inference about the former may be reached based on 
the latter, these involving easier measurements. We point
out that for other populations, and specifically for smaller
fish, the correlation might be smaller and the difference
between the two indices larger. This is reflected in 
the fact that $\E\,X/\E\,Y$ is not a good approximation
to $\E\,(X/Y)$ when the variance of $Y$ is moderate
or big and in particular when $Y$ may be small 
with reasonable probability. 

The HSI is an overall number for the liver quality of 
a stock of fish (in a given year), averaged over subclasses 
of age, sex, and perhaps other identifiable categories, 
like year-class. For larger specimens the female cod 
typically have a higher HSI than the male, for example 
\citep{KrivobokTokareva73, Karlsenetal05}. 
We may represent this as 
\beqn
\HSI=\sum_{{\rm strata}} w(u)\,\HSI(u), 
\eeqn 
with $\HSI(u)$ denoting the index for stratum $u$ and 
$w(u)$ the relative frequency of this stratum among 
all strata. Thus a different way of measuring or 
modelling the overall HSI is via separate modelling 
of the frequencies, say across age and sex, along
with separate modelling or sampling for these groups. 
These considerations also tell us that the HSI number
is a complex quantity that may vary from one year 
to the next via a number of reasons. In particular,
the HSI may change over time not because the individual 
specimens change their liver sizes but because 
the demographical characteristics change 
(e.g.~having a higher proportion of older fish). 

\section*{Statistical modelling of the HSI time series}
\label{section:modelselection}

In the following we will introduce different statistical 
models to study and analyse the bulk 
liver quality index $\HSI=\HSI_\bulk$ of Figure \ref{figure:Hjort}.
A good enough statistical model allows us to answer various 
pertinent questions, predict the future behaviour of the HSI, 
check for anomalies, discover potential outliers, 
and investigate structural changes, break 
points or regime shifts in observed series.
Moreover, a full stochastic model makes it possible 
to study different joint relationships and interactions 
between our main target for inference, the HSI, and 
several explanatory series, like the Kola temperatures, 
the length distribution of the population, 
the mortality rate, and the food supply, as briefly 
pointed to in the introduction above and further 
specified below.

Before we introduce these covariate series we carry out
a separate investigation of the bulk HSI series from 
Figure \ref{figure:Hjort} in itself. 
In order to do so, let $z_{i}$ represent liver quality index 
$\HSI$ for $\year_{i}$ and consider the model where 
\beq
\label{eq:hsi_prototype_model} 
z_{i} = \beta_{0} + \beta_{1} \times \year_{i} 
  +  \eps_i, \quad \textrm{ for } 1859 \le \year_{i} \le 2012, 
\eeq
with $\{\eps_i\}$ taken to be a stationary zero-mean 
Gaussian time series. Typically, the stationary part of 
[\ref{eq:hsi_prototype_model}] 
will be modelled as a low-order autoregressive process.
For a $k$-th order autoregressive model, this means that  
$\eps_i = \rho_1\eps_{i-1}+\cdots+\rho_k\eps_{i-k} + \sigma\delta_i$, 
where the noise terms $\delta_i$ are independent 
and standard normally distributed (some alternative and 
more general types of models will be discussed below). 
Without going into the details, we point out that some 
technical conditions on the $\rho_j$ parameters are 
needed in order to ensure that the resulting model 
is indeed stationary; see \citet[Ch.~3]{Brockwell91}
and \cite{Brillinger75} for a complete and
technical introduction to time series modelling. 

The stationary part of the model [\ref{eq:hsi_prototype_model}] 
introduces lagged dependencies between consecutive years of the HSI 
and is by the simple structure of the autoregressive models 
not difficult to interpret. The residuals $z_t-\beta_0-\beta_t\year_t$ 
depends linearly on the previous values, with the degree 
to which this happens indicated by the size of the coefficients. 
We also take the opportunity to point out that the class 
of autoregressive models is able to approximate 
any type of stationary dependency structure.  

In model [\ref{eq:hsi_prototype_model}] we have also included 
a linear drift or trend, and as mentioned more complex 
(and perhaps more realistic) relationships including 
covariates will be discussed later. 
We will question whether $\beta_1$ is zero or not. 
If $\beta_1$ is significantly smaller than zero, for example, 
it might have important implications for future behaviour 
and the general understanding of the liver quality index,
and indeed of the fish population itself. 

\begin{figure}[ht]
\includegraphics[width=0.45\textwidth]{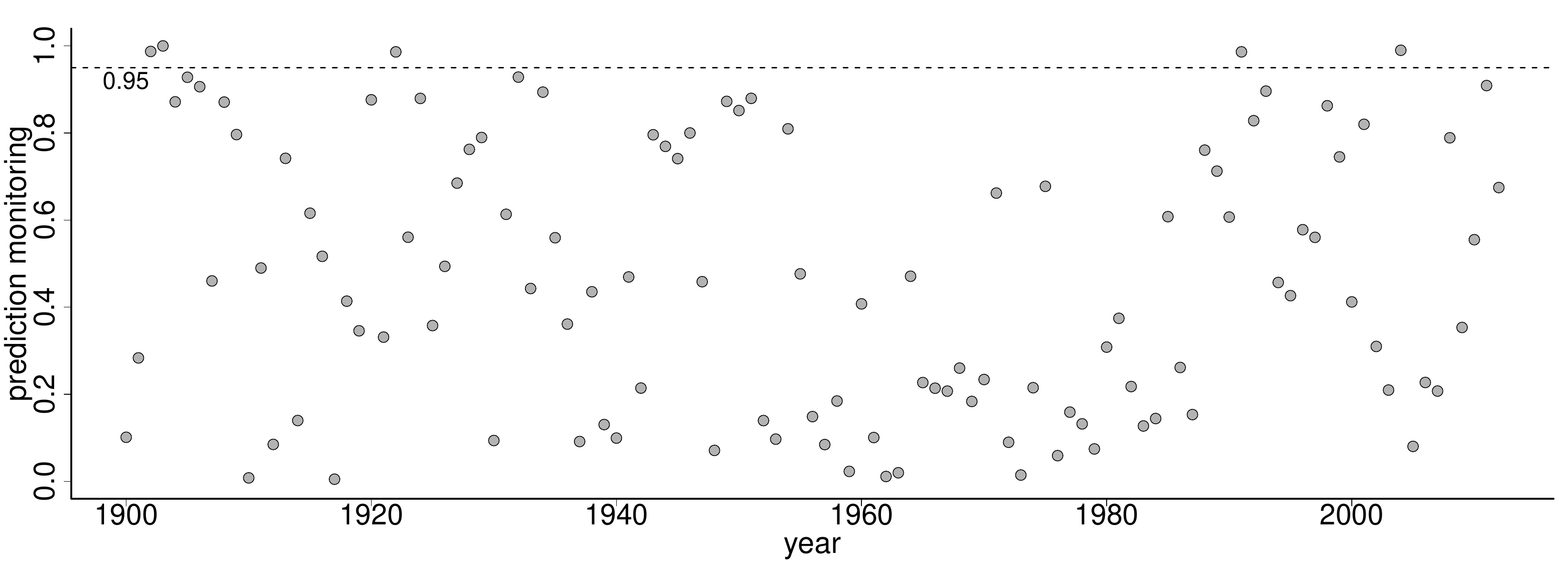}
\caption{
Sequential one-year-ahead predictions (1900--2012) from 
an autoregressive model of order two, translated to 
prediction monitoring values $m_t=\Gamma_1(d_t^2)$
(see text). The plot suggests the overall quality
of the prediction model is good.}
\label{figure:PredictedHSIValues1}
\end{figure}

\begin{figure}[ht]
\centering 
\includegraphics[width=0.30\textwidth]{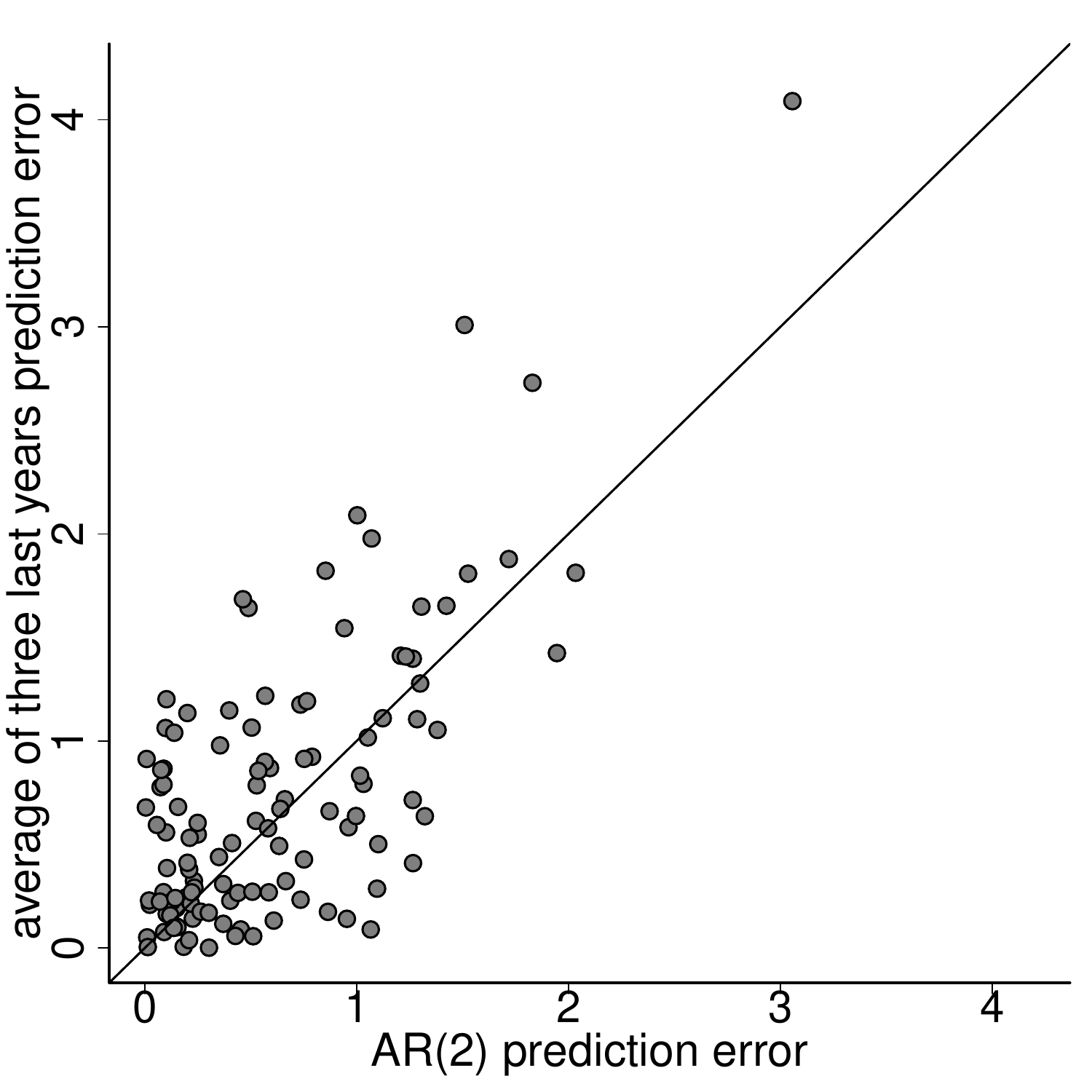}
\caption{
Sequential one-year-ahead predictions (1900--2012) 
from an autoregressive model of order two 
(as in Figure \ref{figure:PredictedHSIValues1}),
compared to the predictions obtained by using the average of 
the HSI of the three preceding years, as a predictor. 
On average, over the period of 112 years, using the 
autoregressive model results an average absolute error of size 0.61, 
while the more naive average of the preceding years gives 
an absolute error of 0.75. This illustrates the potential 
gain by building more complex statistical models.}
\label{figure:PredictedHSIValues2}
\end{figure}

Figures \ref{figure:PredictedHSIValues1} and \ref{figure:PredictedHSIValues2}
shed light on the ability of the predictive ability
of model [\ref{eq:hsi_prototype_model}], for the case
of one-year-ahead predictions for the HSI series, using
an autoregressive model of order two. 
The first plot provides prediction monitoring values 
$m_t=\Gamma_1(d_t^2)$, where $d_t=(z_t-\hatt z_t)/\hatt\tau_t$ 
is the standardised prediction error made by computing
$\hatt z_t$ to predict the actually observed $z_t$,
with calculated prediction error $\hatt\tau_t$; also,
$\Gamma_1$ is the distribution function of a chi-squared 
variable with one degree of freedom. The idea is that 
if the model used for prediction is good, the $d_t$
numbers will be close to standard normal, which means
that the $m_t$ values will be close to uniformly distributed
on the unit interval. If the model does not fit well,
the $m_t$ values will tend to be closer to one. 

\subsection*{Is the HSI time series stationary?} 


Analysing the complete $\HSI_\bulk$ time series 
of Figure \ref{figure:Hjort} using standard tests for 
stationarity, there is little evidence suggesting 
that the HSI does not satisfy the conditions for 
being stationary. This depends however to some extent 
on the time window considered. A Dickey--Fuller type test 
\citep{DickeyFuller79}, for example, used on the full HSI series
1859--2012, rejects the null hypothesis (with a p-value less 
than 0.01) that the series has a unit root. 
Along with graphical diagnostics this provides evidence 
that the Hjort series is well modelled via 
e.g.~low order autoregressive models. 
By making sequential AIC analyses, in the same spirit as 
with those reported on in Figures \ref{figure:aic_race} 
and \ref{figure:Kola:WinterVSAnnual}, however, 
we get a less conclusive picture and observe that 
the linear effect of including $\beta_{\year} \times \year_{i}$ 
in the model has significant importance, for several 
long periods of time. We will therefore keep 
the linear component in our baseline model for now;  
this also serves as a minimal model to attempt to 
improve upon when we introduce more complex models 
and covariates in later sections. 

\begin{figure}[ht]
\centering
\includegraphics[width=0.45\textwidth]{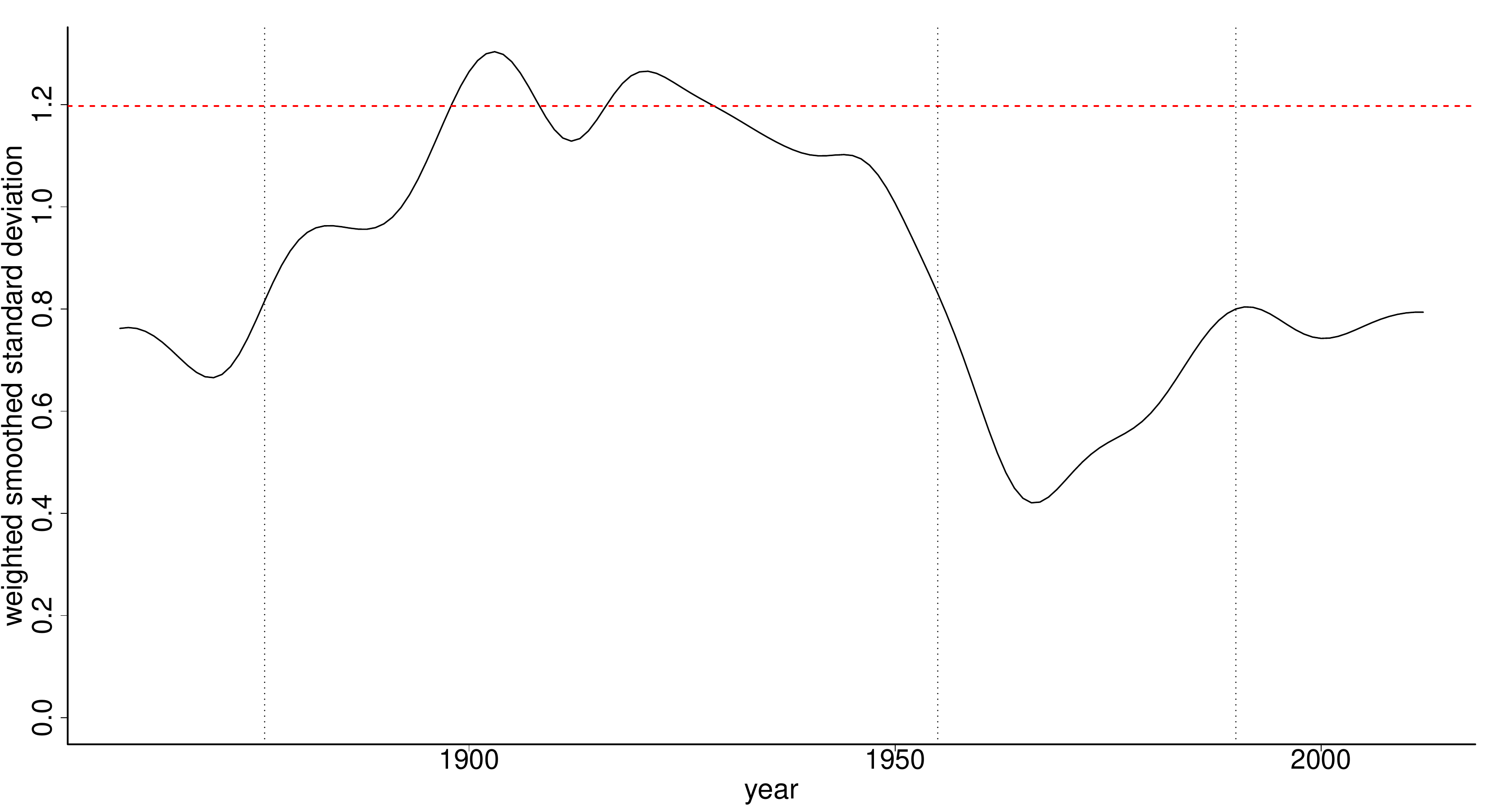}
\caption{Estimated standard deviation for the HSI bulk series, 
demonstrating a varying level of variability over time. 
The main drop in variability is around 1955 (see the text).
The vertical lines are for 1876 (the start of the observed HSI index), 
1955, and 1990. The horizontal line indicates the overall 
standard deviation estimate using the complete series.}
\label{figure:HjortWeightedStandardDeviation}
\end{figure}

While the brief considerations above concerned the 
stationarity or not of the mean function, a different
aspect of the HSI series is its variability level
around the mean function. 
Visual inspection of the series might indicate 
a non-constant level of variability. This is borne
out of careful estimation and testing procedures,
indicating a certain decrease in variability level 
during the years 1955--1990. The standard deviation parameter 
examined now, say $\sigma_t$ at year $t$, 
is that associated with the distribution of $\HSI_\bulk$ 
over time, not that of the precision of an individual
data point. The drop in standard deviation appears 
to stem from around 1955 
(Figure \ref{figure:HjortWeightedStandardDeviation}),
at a time when trawling becomes the dominant catch method 
in Norwegian fisherie. 
This drop in $\sigma_t$ is also associated with 
generally higher exploitation rates up to recent times 
\citep{Jorgensen90, ICES2014, Kjesbuetal14A}, 
causing an abrupt fall in overall body size
\citep{Jorgensen90, Kjesbuetal14B} 
and thereby in HSI which is positively size dependent 
\citet{Kjesbuetal14B}.
The same period shows up in Figure \ref{figure:PredictedHSIValues1}, 
\ref{figure:HSILoglikMax} as moderate irregularities, 
and is also reflected in models we have invesigated
of the time-varying coefficients type
(see our concluding remark D below). 


For the illustrative purpose of the present section 
the deviation is not too severe, and approximating 
the standard deviation function with a constant 
will not become statistically troublesome, for most purposes
(see the appropriate FIC discussion below).

\subsection*{Structural changes} 

The sea is big and nature is sometimes changing her ways.
There are occasions where a certain statistical model 
is in force for a certain number of years, after which 
the underlying parameters change significantly and 
perhaps rapidly, pushing certain aspects of interest
into a new state of equilibrium. This has e.g.~arguably 
happened regarding the ways in which the Kola winter temperature
and the HSI series and their interplay have developed. 

There are various statistical techniques devoted to
studying such phenomena, from testing the null hypothesis
that no significant change has taken place during 
a certain time period of time to estimating the position
of a break point in case such a discontinuity has 
taken place; see \citet{FrigessiHjort02} for a general
discussion and overview and \cite{HjortKoning02} 
for a class of such methods. Here we briefly outline 
one particular method, associated with a certain 
graphical plot for checking constancy and looking for
break points. Assume a certain model is put to work,
involving a parameter vector 
$\theta=(\theta_1,\ldots,\theta_p)$ of length $p$, 
leading in particular to the model-based log-likelihood 
function $\ell_j(\theta)$ associated with years $1,\ldots,j$. 
Thus we may for each time period $1$ to $j$ compute 
the maximum likelihood estimate $\hatt\theta_j$ 
and associated log-likelihood maximum value $\ell_{\max,j}$, 
say, as long as $j\ge p$. From the full sequence
of observations, over years $1,\ldots,n$, we may thus
monitor both how the parameter estimates and the 
log-likelihood maxima develop over time. One particular
monitoring bridge function is then 
\beq
\label{eq:bridge}
B_{n,j}=\rootn\{n^{-1}\ell_{\max,j}-(j/n)\hatt a\}/\hatt\kappa 
   \quad {\rm for\ }j=p,\ldots,n, 
\eeq 
where $\hatt a=n^{-1}\ell_{\max,n}$ and $\hatt\kappa$,
the latter an estimate of $1/\rootn$ times the 
standard deviation of $\ell_n(\theta)$, are both based 
on the full data set. Note that the process ends in $B_{n,n}=0$.
The point now is that 
(i) if the underlying model in fact
has not changed, after all, then the $B_{n,j}$ process
[\ref{eq:bridge}] behaves as a so-called Brownian bridge,
with controlled fluctuations; and 
(ii) if there is a break point, the plot will help 
in identifying its position. 

\begin{figure}[ht]
\includegraphics[width=0.47\textwidth]{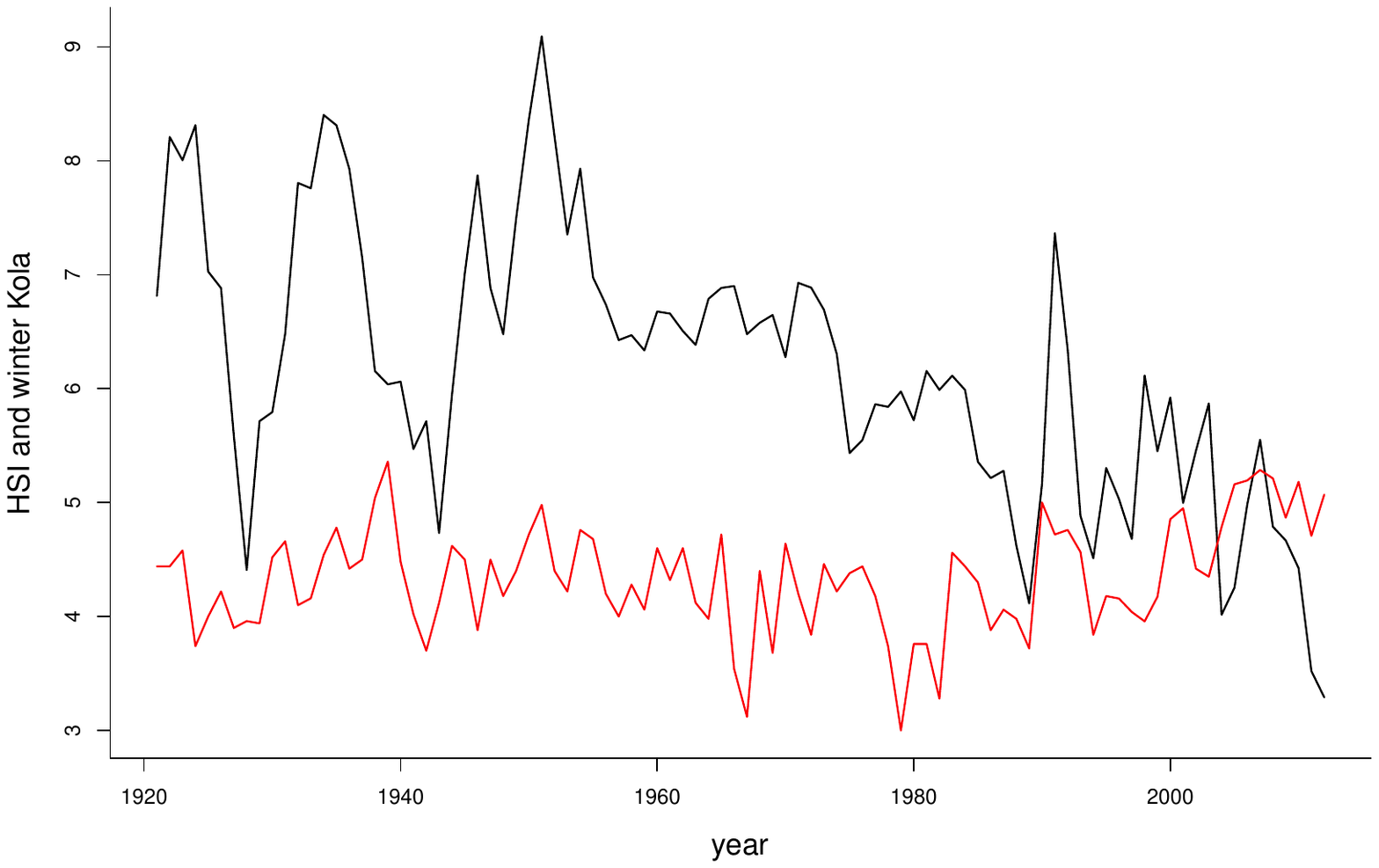}
\caption{The HSI series for 1921--2012 (black), 
along with average Kola winter temperature (red, in degrees Celsius).}
\label{figure:HjortWinterKola}
\end{figure}

\begin{figure}[ht]
\includegraphics[width=0.47\textwidth]{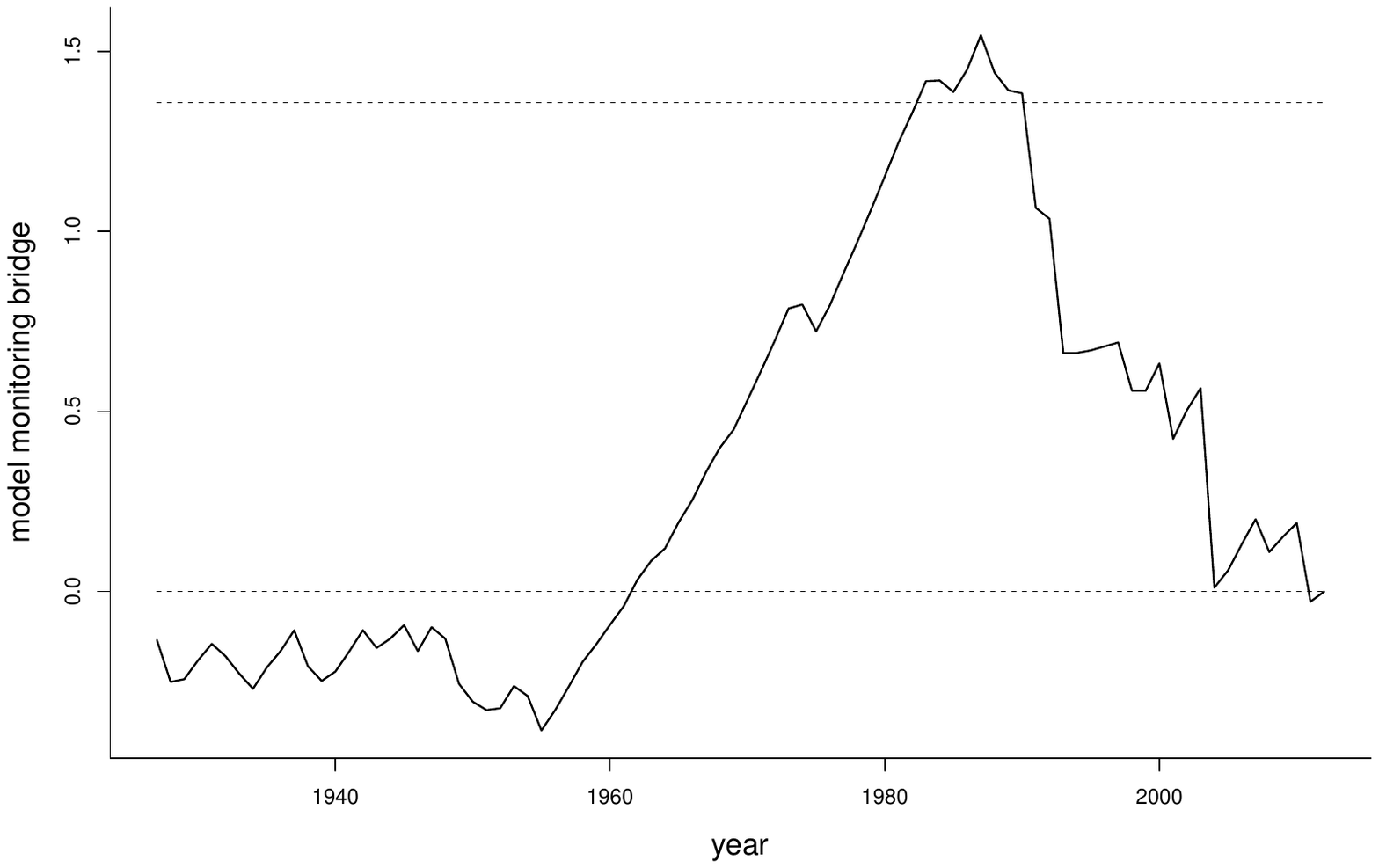}
\caption{Monitoring bridge plot for the model 
$\HSI_i=\beta_0+\beta_1x_{i-1}+\sigma\eps_i$ for the $\HSI_\bulk$ series, 
with $x_{i-1}$ the previous year's average winter Kola temperature 
and the $\eps_i$ a standardised AR(1) process. The plot suggests 
there is a regime shift around year 1990, with model
parameters taking on new values.}
\label{figure:HSILoglikMax}
\end{figure}

Figure \ref{figure:HSILoglikMax} provides an application 
of this method, associated with the autoregressive model 
\beqn
\HSI_i=\beta_0+\beta_1x_{i-1}+\sigma\eps_i 
   \quad {\rm for\ }1921{\rm\ to\ }2012,
\eeqn 
where $x_{i-1}$ is the average Kola winter temperature 
from the year preceding $\HSI_i$ (cf.~further discussion
below, in the section on covarying factors and further models),
and the $\eps_i$ for this illustration follows an 
autoregressive model of order one.  
The two series are shown in Figure \ref{figure:HjortWinterKola}.  
A Brownian bridge stays within $\pm1.358$ with probability 
0.95, so any $B_{n,j}$ values observed outside this band indicate
that the underlying model has not stayed constant over 
the time window considered (corresponding to testing
the hypothesis of a constant model with significance 
level 0.05). Here the maximum value is indeed higher 
than 1.358, as indicated in the figure. 
The data hence suggest there is a break point for the model 
around year 1990, e.g.~with the model switching parameter 
values around that time.  

\section*{Model selection}
\label{section:ModelSelection}

Selecting an appropriate model is an important and 
integrated part of the statistical inference process. 
In many or most situations the statistician 
will have more than one reasonable candidate for modelling 
the phenomena under study. 
Selecting an appropriate model for the final report and 
analysis is therefore of significant importance, 
hence necessitating the notion of and research field 
of statistical model selection.

Although not explicitly pointed to, parts of the above 
discussion regarding the validity of the 
model and reliability of the underlying assumptions, 
e.g.~checking for non-stationarity in the HSI series
and examining potential trends, are actually questions 
related to model selection.
Selecting the `best' model among a set of potential and 
reasonable candidates has an ongoing and long history, 
with techniques ranging from visual inspection, 
goodness-of-fit tesing and so-called model information criteria; 
for a general introduction to these themes see \cite{ClaeskensHjort08}.

Models with many parameters may become ineffective 
due to estimation variability, whereas slimmer models 
with fewer parameters might suffer from modelling bias. 
A `good' model selection strategy should balance out
complexity against simplicity and precision in a reasonable way. 
Models should be as simple as possible, but not simpler,
as Einstein implied. The preferred model should be rich enough 
to capture the essential features, and with high enough precision 
to be useful, and at the same time still be simple enough 
to be comprehensible and possible to handle regarding
fitting and inference techniques. Different model selectors
balance the desiderata of `low variance' and `low bias'
in different ways. 

\subsection*{The AIC, BIC and other information criteria}
\label{section:IC}

Among the more popular model selection strategies 
are Aka\-ike's information criterion (AIC; \cite{Akaike73}), 
the Bayesian information criterion (BIC; \cite{Schwarz78}) 
and the focused information criterion 
(FIC; \cite{ClaeskensHjort03}, see below). 
These have  considerable appeal, since they are typically simple 
in both structure and use, resulting in model scores which 
can be used to rank candidate models from best to worst 
in accordance with a well-defined measure of discrepancy. 
The practical simplicity of AIC and BIC has perhaps 
led to uncritical use. This is especially true for the AIC, 
which is often used without any concern for the 
underlying motivation; see \cite{Hermansen14b}.

The AIC is defined as 
\beqn
\AIC = 2\,\textrm{log-lik}_{\max} - 2 p, 
\eeqn 
with $\textrm{log-lik}_{\max}$ the maximal value of the 
log-likelihood function and $p$ the number of parameters used 
in the model. The BIC has a similar structure, 
viz.~$\BIC = 2\,\textrm{log-lik}_{\max} - p\log n$, 
but stems from a quite different motivation, to be 
commented on in a moment. Both lead to one `best' model, 
aiming respectively at the one minimising a certain 
Kullback--Leibler divergence from the underlying true data 
generating mechanism to the model in question, 
and the one maximising the posterior model probability
in a Bayesian framework. 
These are global perspectives, preferring models that 
aim at capturing the main characteristics 
of the underlying data generating process. 

To illustrate the use of these criteria, consider modelling 
the HSI series as $z_t=\beta+\sigma\eps_t$, with the $\eps_t$
forming a zero-mean standardised normal autoregressive process 
$\AR(k)$ of orders 0 (corresponding to independence), 1, 2, 3, 4, 5. 
The parameter dimensions of these models are
respectively 3, 4, 5, 6, 7 (since $\beta$ and $\sigma$
are parameters to be estimated for each of them). 
Table \ref{table:AICandBICscores} gives the AIC and BIC scores,
yielding in particular two different advices. 


\begin{table*}[ht]
\centering
\begin{tabular}{c|cccccc}
$k$ & 0 & 1 & 2 & 3 & 4 & 5  \\
\hline
 dim & 2 & 3 & 4 & 5 & 6 & 7 \\
 AIC  &  $-495.5$ & $-363.7\,$   & $-363.5^{\ast}$ & $-365.0$ & $-367.0$  & $-368.9$  \\
 BIC  &  $-501.6$ & $-372.9^{\ast}$ & $-375.6\,$       & $-380.2$ & $-385.2$ & $-390.2$  \\
\end{tabular}
\caption{
AIC and BIC scores when fitting stationary autoregressive models
of orders 0, 1, 2, 3, 4, 5 to the HSI bulk index time series of 
Figure \ref{figure:Hjort}, with the best models indexed with 
an asterix. The order zero model, corresponding to independence, 
is judged too simple by both criteria. The BIC has the $\AR(1)$ as its 
winner, where the AIC on the other hand scores suggests $\AR(1)$ 
and $\AR(2)$ are about equally good, with a slight preference 
for the latter.}
\label{table:AICandBICscores}
\end{table*}

\begin{figure}[ht]
\includegraphics[width=0.47\textwidth]{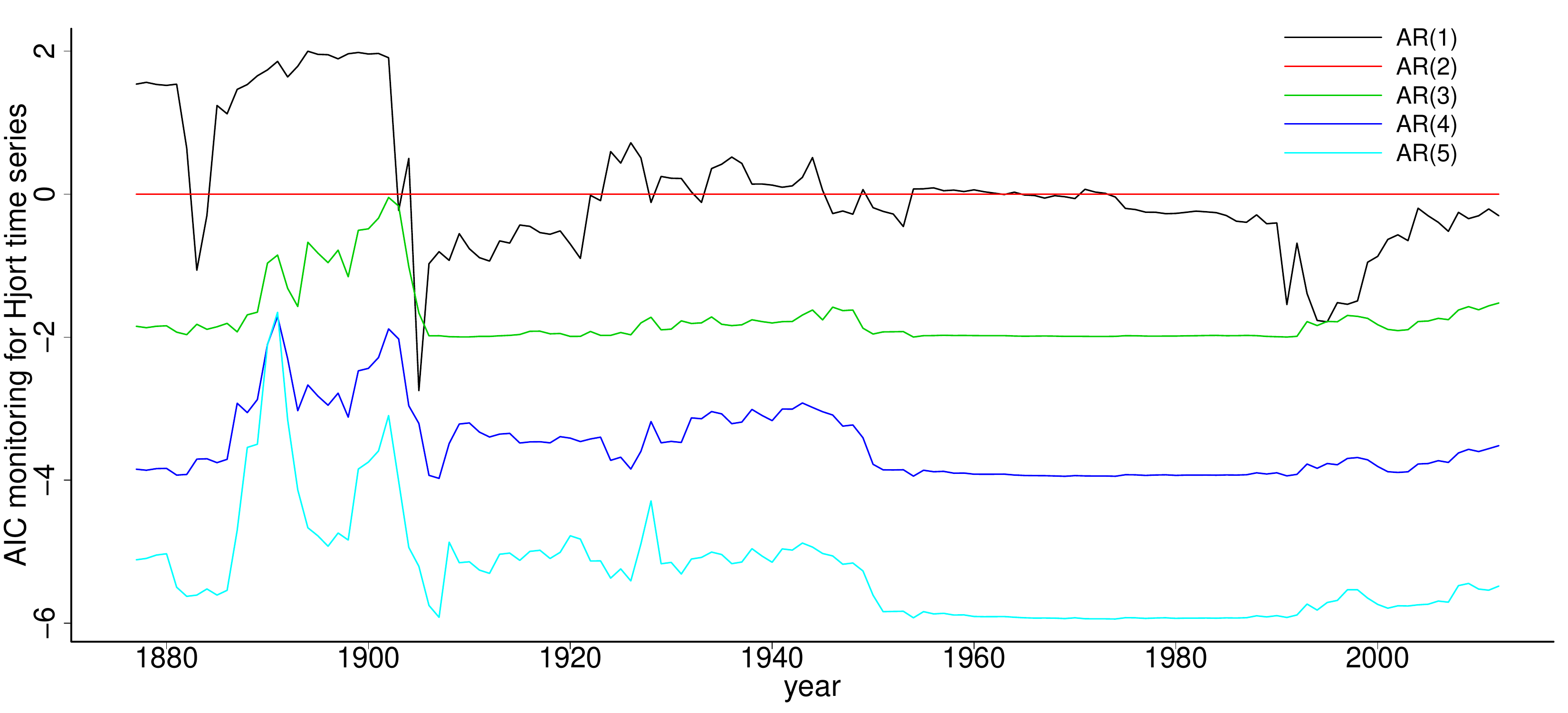}
\caption{
Sequential AIC score differences, relative to the autoregressive 
model of order two. High values indicates model fit 
better than with the $\AR(2)$. 
Higher order models (of order 3 or more) behave more 
or less the same in terms of model choice quality.} 
\label{figure:aic_race}
\end{figure}

Another illustration is the dynamic plot of 
Figure \ref{figure:aic_race}, showing relative AIC scores 
for models $\AR(1)$ to $\AR(5)$ as these progress 
with more data accumulated over time. The relative
AIC score in question is $\AIC_{\AR(k)}-\AIC_{\AR(2)}$.
The plot indicates first that higher order models
do not contribute significantly, and secondly that 
something noteworthy takes place around 1903, with
a drop in exploratory power. This is associated
with the all-time record low value for the HSI in that year. 

\subsection*{The FIC}
\label{section:FIC}

Instead of aiming at a model that is `reasonably good at everything', 
the motivation underlying the FIC is that the intended 
use of the model and the focus of the investigation should 
play a central part of the selection procedure. Only 
rarely is one and the same model good for all purposes.  
This is e.g.~evident for regression models, where some 
covariates may be important for some types of questions 
but of lesser importance for other aspects of what is being studied. 
Compared to other classical information criteria, the FIC allows 
the precise intention of the analysis to be taken into account 
when selecting the model. The FIC sidesteps the often 
unachievable goal of finding one `correct' model for 
all uses and aims instead at finding the model which 
is best suited for answering focused questions, 
one focus at a time. 

The FIC was introduced in \cite{HjortClaeskens03} and 
\cite{ClaeskensHjort03} and is based on estimating and 
comparing the accuracy of individual model-based estimators 
for a chosen focus parameter, which we here denote $\mu$. 
The focus $\mu$ ought to have a clear statistical interpretation 
across candidate models. For a given candidate model, 
$\mu$ is then expressed as a function of this model's parameters. 
For the model  [\ref{eq:hsi_prototype_model}], the parameters 
are $\beta_{0}, \beta_{1}, \sigma$, along with the autoregressive 
parameters $\rho_{1}, \ldots, \rho_{k}$ specifying the dependency structure. 
Valid and relevant focus parameters include quantiles, 
regression coefficients, a specified lagged correlation, 
and various types of predictions and data dependent functions 
like the probability that the future HSI index will be below 
a given threshold (say the 3.0 value reached in the year 1903), 
given the observed history of liver quality values. 
See \cite{HermansenHjort15b} for details pertaining to 
FIC methods for time series models.  

Suppose there are candidate models $M_1,\ldots,M_k$,
leading to focus parameter estimates $\hatt\mu_1,\ldots,\hatt\mu_k$,
respectively. The underlying idea leading to the FIC 
is to estimate the mean squared error (mse) of $\hatt\mu_j$ 
for each candidate model and prefer the model that achieves 
the smallest value. The mean squared error in question is 
\beq
\label{eq:mse}
r_j=\E\,(\hatt\mu_j - \mu_{\true})^{2} 
   = \Var\,\hatt\mu_j 
  + \textrm{bias}(\hatt\mu_j)^{2},
\eeq
comprising the variance and the squared bias 
in relation to the true parameter value $\mu_\true$. 
Thus the FIC consists of finding ways of assessing,
approximating and then estimating the $r_j$ for 
each candidate model, and the winning model is the
one with smallest $\hatt r_j$. How this may be done depends
on both the candidate models and the focus parameter, 
as well as on other characteristics of the underlying situation.
The FIC apparatus hence leads to different types of formulae 
in different setups; 
see \citet[Ch.~5 \& 6]{ClaeskensHjort08} for a fuller
discussion, illustrations, and generalisations.
For time series models, as met when modelling the HSI
series and its covariate series, certain complexities 
are involved, see \cite{Hermansen14b}. Importantly,
the FIC may lead to different models being pinpointed 
as best, for different foci. 


\begin{figure*}[ht]
\centering
\includegraphics[width=0.95\textwidth]{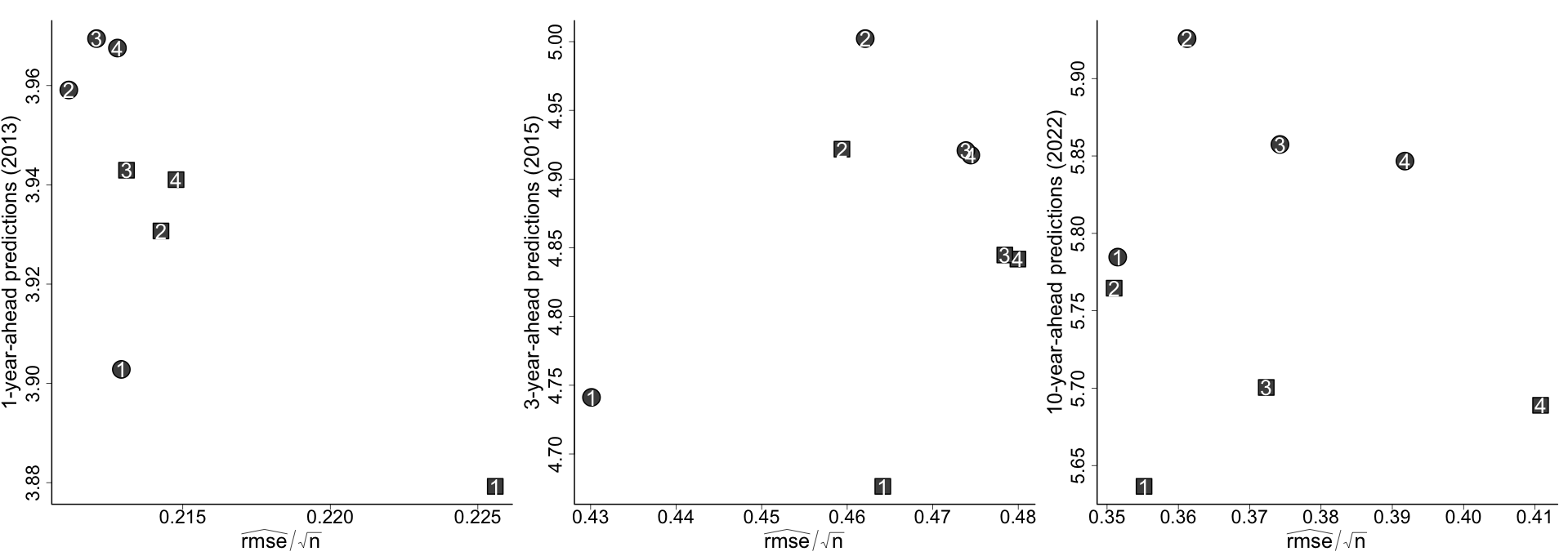}
\caption{
FIC plots for the predicted liver quality index, 1, 3, 10 years from now.
The models are autoregressive models of order 1--4 (indicated by the
number inside each point) with a linear trend (circle) and also 
without (square). The AIC and BIC, which do not differentiate between 
the different foci, both prefer the autoregressive model of order 2
without the linear trend.}
\label{figure:FIC:HSI_with_time}
\end{figure*}

\begin{figure*}[ht]
\centering
\includegraphics[width=0.95\textwidth]{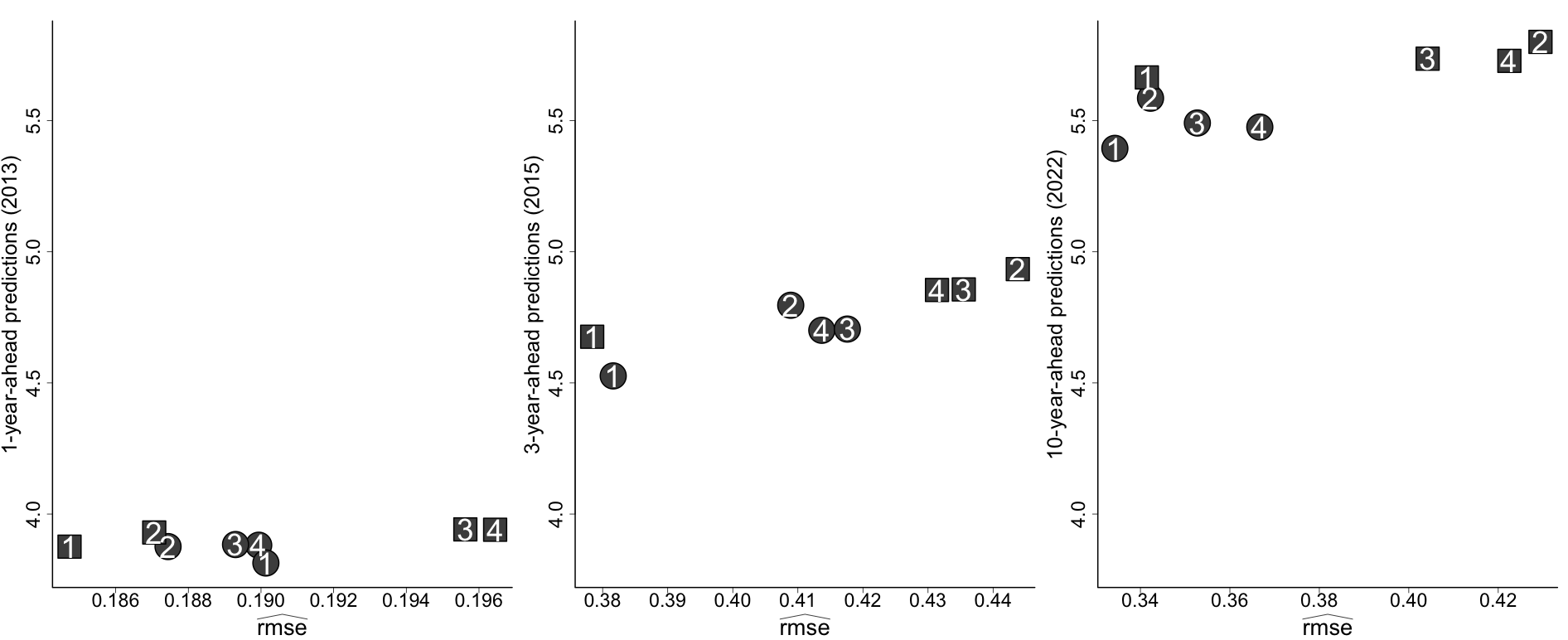}
\caption{
FIC plots for the predicted liver quality index, 1, 3, 10 years 
from now. The models are autoregressive models of order 1--4 
(indicated by the number inside each point) with and without 
(square) a linear trend (circle). z
The AIC and BIC, which do not differentiate between the 
different foci, prefers respectively the autoregressive model 
of order 1 and 2 both without the linear trend.}
\label{figure:FIC:HSI_with_time}
\end{figure*}

We will now apply the FIC strategy for the HSI study.
We consider eight natural candidate models, 
corresponding to autoregressive order 1, 2, 3, 4,
with or without a linear trend component over time.
The focus function we consider first is 
\beqn
\mu_{\textrm{pred}}(\year) = Z_{\year}, \quad \textrm{ for } \year > 2012, 
\eeqn 
and more specifically we aim for this illustration 
at models that are good at predicting 1, 3 and 10 years 
into the future (i.e.~2013, 2015, 2022, as seen from year 2012, 
the current endpoint of the HSI bulk series). Carrying out the 
FIC step aids in understanding the short, medium and longer-term 
mechanisms involved in the HSI process. 
The results are shown in Figure \ref{figure:FIC:HSI_with_time}.
These FIC plots show predictions on the y-axis and 
FIC scores, i.e.~estimated root mean squared errors 
$\hatt r_j^{1/2}$, on the x-axis.  

For the one-year ahead predicted HSI value of 2013, 
we observe that essentially all models 
are in more or less complete agreement (apart from the 
too simple model which uses independent errors; see below). 
This is also reflected by the small differences 
in the estimated root-mean-squared error values. 
For such large sample sizes ($n = 154$) and such relatively 
simple focus questions, we do not expect the quality and the 
estimates to deviate considerable (as long as 
all fitted models are all fairly reasonable).

It is worth pointing out that the FIC judges the 
autoregressive model with linear trend to be the best 
model for predicting the liver quality ten years ahead 
(here 2022). This is in contrast to the observations 
made earlier that the model with linear trend 
seemed to be inferior to the simpler stationary model with 
a constant trend. As a general observation, we see 
that as we try to predict further and further into 
the future, the models with a linear trend
start to dominate among the `best’ models. This 
suggests that there might be a (significant) linear 
effect needed to explain the long term
behaviour of the HSI index. 

The models with independent errors (referred to also as 
autoregressive models of order zero) are not included in the 
plot. These resulted in estimated root-mean-squared 
errors of more than 1.5, falling outside the natural
scale spanned by the other and better models 
in Figure \ref{figure:FIC:HSI_with_time}. 

There is even more room and need for FIC type model
building and selection tools when covariate information
is taken into account, as we shall see in the following
section. With covariates on board the number of 
natural candidate models also increases rapidly.  



\section*{Covarying factors and further models}
\label{section:covariates_and_other_models}

In this section we briefly report on investigations 
on whether certain factors may be seen to be significantly 
correlated with the HSI series. In cases where there
is such an identifiable correlation, further considerations
and analyses might be called for when it comes to 
determining what causes what. 
The explanatory variables we consider are 
\begin{enumerate}
\item[(i)] Kola temperatures 
(annual and winter), with data provided by PINRO (www.pinro.ru/index-e.htm); 
\item[(ii)] average length distribution, with data from the 
IMR long-term catch sampling programme, 
see details in \citet{Kjesbuetal14B}; 
\item[(iii)] fish mortality rate F (the usual parameter associated 
with continuous fishing and natural mortality, 
see e.g.~\citet[Ch.~10.3]{HilbornWalters92}), 
with data from \cite{ICES2014}; and 
\item[(iv)] an index for the amount of food available 
\citep{ICES2014}, the latter defined as the ratio 
of biomass for caplin with spawning stock biomass 
for the northeast Arctic cod \citep{Kjesbuetal98}. 
\end{enumerate} 
As a curiosum, since \cite[p.~186]{JHjort14} found it 
necessary to briefly dismiss the hypothesis apparently 
put forward by \citet[Ch.~VII.3]{HellandHansenNansen1909} 
that the annual sunspot numbers could influence the liver
quality of cod, we will also compare the sunspots series
with the $\HSI_\bulk$.  
In addition to studying each of these 
and their respective connection to the Hjort series 
on their own, we will also use the forthcoming discussion 
to select some of the covariate series for a further 
combined analysis. 

\subsection*{(i) Kola temperatures}
\label{section:Kola}

In \cite{Kjesbuetal14B} the connection between the annual 
Kola temperature (1900--2012) and the $\HSI_\bulk$ 
was studied. The two time series exhibit a covarying  
pattern lasting for many years, but the apparently 
strong relationship seems to have ended somewhere after 1960.
We commented on this in connection with Figures 
\ref{figure:HjortWinterKola} and \ref{figure:HSILoglikMax}. 
In the following section we will continue this discussion 
and also try to unveil where the separation takes place.

Instead of working with the annual average temperature 
we will use what we define as the average winter 
temperature, averaging the monthly means from 
start of October (previous year) to start of March (current year). 
The data used are monthly averages of Kola temperatures 
from 1921--2012. These winter months may carry more 
relevant information since it is during this period the cod 
recuperates for spawning. 

\begin{figure}[ht]
\includegraphics[width=0.48\textwidth]{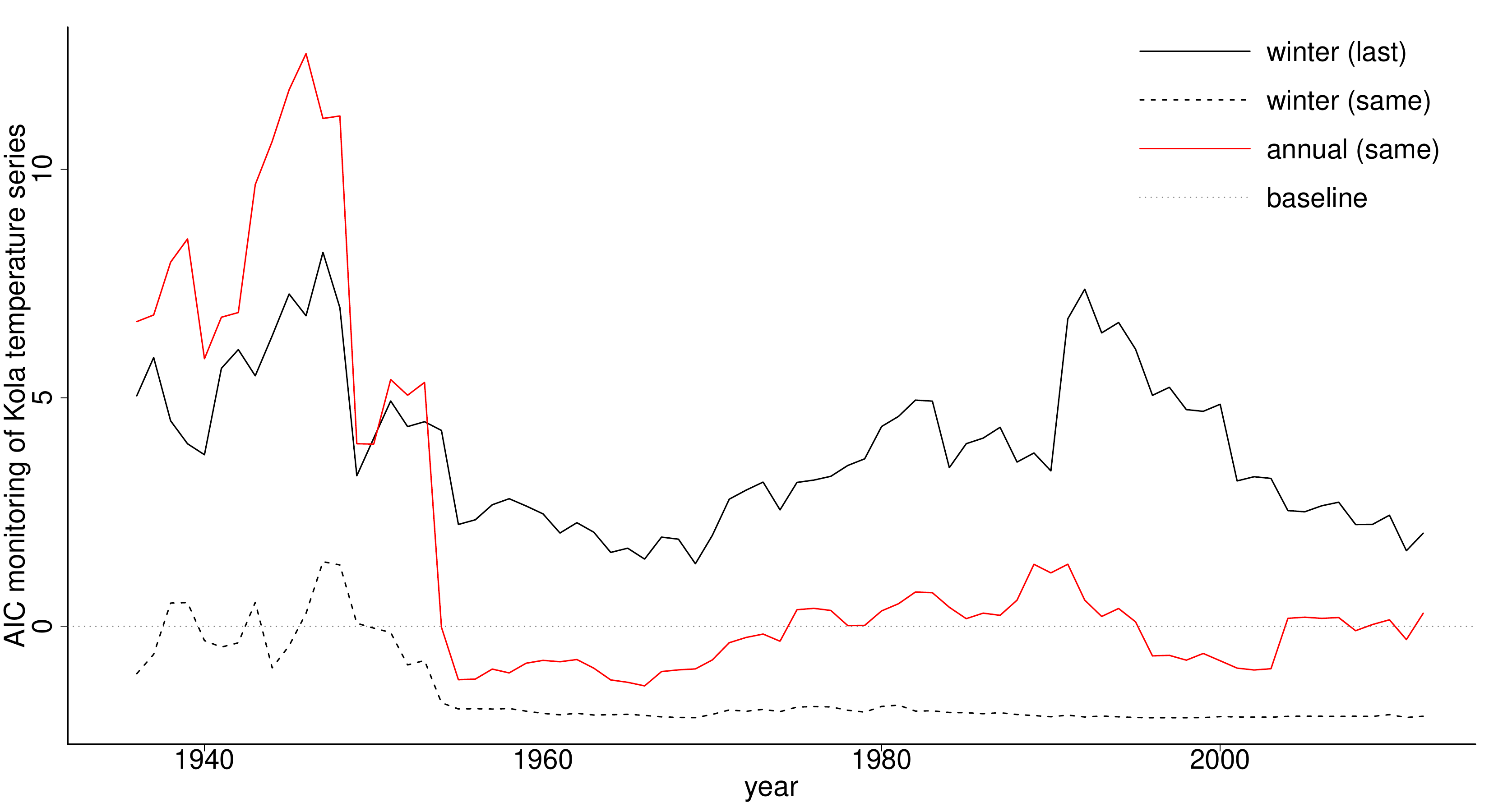}
\caption{
Sequential AIC score difference from the baseline model 
with an autoregressive model of order two. 
Using the average winter Kola temperature is seen 
to systematically improve the model also after the years 
with strong correlation (prior to 1960).}
\label{figure:Kola:WinterVSAnnual}
\end{figure}

By studying the effect of taking different
time lags, i.e.~the relationship between the HSI index 
for a given year and various combinations of preceding 
predictors (in this case the average Kola winter temperature), 
we observe that the average winter temperature 
of the previous winter carries more relevant information 
and also provides a more significance signal than the 
effect from the current period, 
see Figure \ref{figure:Kola:WinterVSAnnual} for details.
This is not evident if the raw correlations are used 
as a measure of information `quality', however, where 
we observe that the raw correlation between last 
Kola winter temperature and $\HSI_\bulk$ is $-0.02$ 
and for the last annual temperature is $-0.03$. 
The corresponding correlation estimates with the current 
year are $-0.07$ and $-0.15$, which might suggest 
that the annual temperature of the same year is 
equally relevant; these correlations are all on the 
tiny side of the spectrum, however.  

\subsection*{(ii) Length index}

\begin{figure}[ht]
\includegraphics[width=0.48\textwidth]{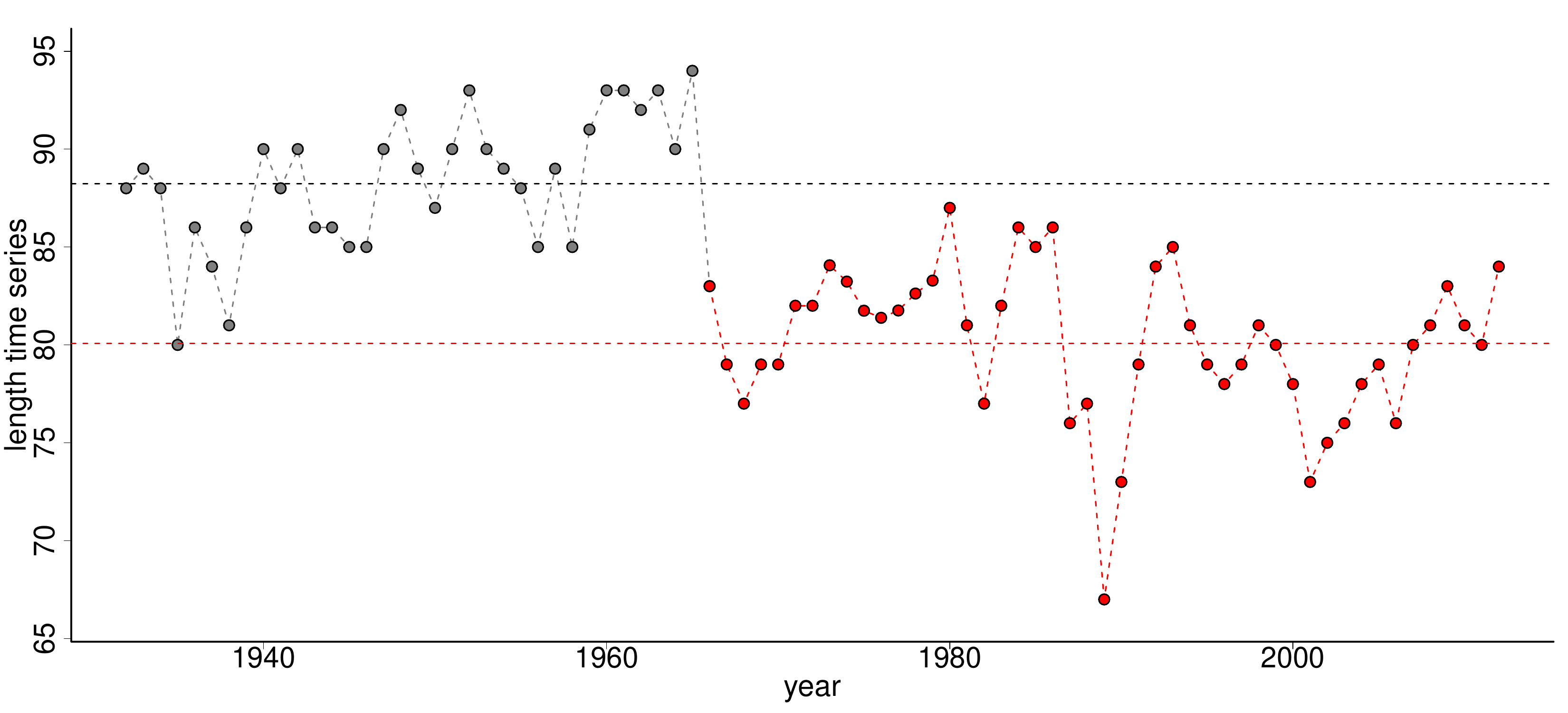}
\caption{
The average length time series with reconstructed length values 
for the period 1973--1979. The observed data 
suggest that there might be a jump discontinuity around
the mid 1960ies. The dotted lines show the estimated 
mean signal before and after the potential jump.}
\label{figure:Length:jump}
\end{figure}

The average length series of Atlantic cod (1932--2012) 
is shown in Figure \ref{figure:Length:jump}. The raw correlation 
between the HSI index and the length series is about 0.51, 
which is quite high, and in fact the largest among the explanatory 
variables considered here. This also suggests that it 
should be a good predictor, since the correlation 
is essentially a measure of linear relationship. Also, 
by undergoing a similar analysis as presented in 
Figure \ref{figure:Kola:WinterVSAnnual}
we observe that as a predictor for HSI, the length average 
for the same year carries the most information about 
the liver quality index. This is in contrast to 
the other series discussed here, where all are seen 
to carry the most information about the current status 
of the HSI if the previous year is used as a explanatory 
variable.  

There is growing evidence that fluctuations in expressions 
of body condition (e.g.~HSI) and reproductive investment 
are not only dependent upon the current environmental 
situation, but that the influence also dates further back 
in time, in line with statements in \cite{Stearns92}. 
This should in particular be true for the present capital 
breeding Northeast Arctic cod undertaking long spawning 
migration, which may show high levels of omission of 
spawning (`skipping') \citep{Skjaeraasenetal12}. 

Unfortunately, there is a notable gap of missing 
values (1973--1979) in the length series, which renders 
complete joint analysis impossible (we can not use 
observations prior to 1980). To overcome this, we 
reconstruct the missing values using the same  
methodology presented in the discussion on combination 
of information; see the discussion below. 

The observed length series (Figure \ref{figure:Length:jump}) 
indicates that there might be a change in the underlying 
model around 1965. 
From a statistical model building perspective discontinuities 
or change points in the underlying model may have critical effects 
on the final analysis, inference and the validity 
of the conclusions made. Here, the effect of the jump 
is perhaps further highlighted by the fact that there 
seems to be an increasing trend prior to the sudden 
change in around 1965. There are various tools constructed 
to find and work with models that have jump discontinuities; 
see e.g.~the jump information criterion JIC of \cite{GHH15}. 
Since we are not studying the length series on its own, 
but are mostly interested in the effect on the liver 
quality index and are therefore merely using it as a covariate, 
the potential discontinuity point (also compared 
to the overall variation) will not violate 
the underlying assumptions to an extent necessary 
for introducing more complex modelling tools. 

\subsection*{(iii) Mortality rate}

\begin{figure}[ht]
\includegraphics[width=0.48\textwidth]{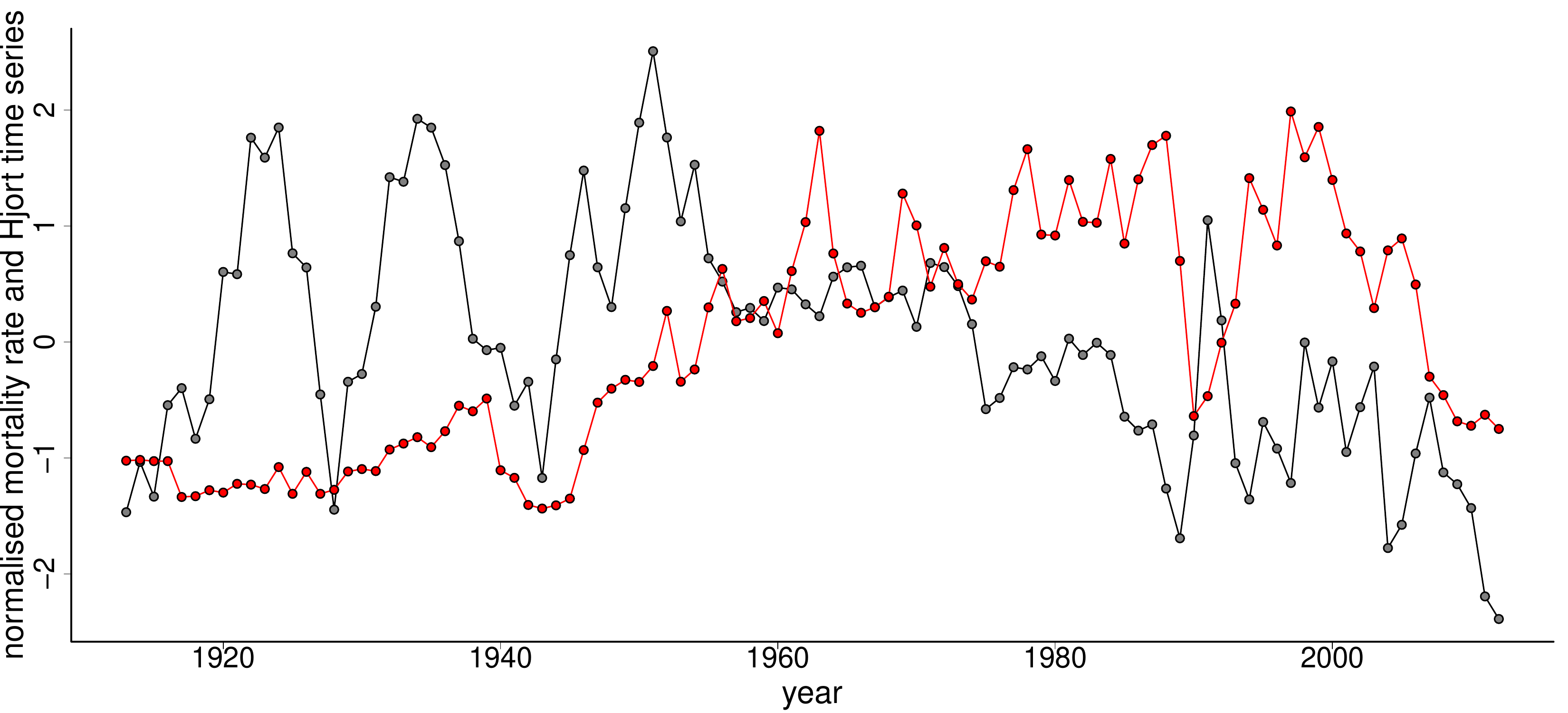}
\caption{
The mortality rate series (red) along with the HSI bulk
series (black). The long persistence 
in the series and almost linear relationship over time 
may be signs of potential long-range dependencies.}
\label{figure:F:data}
\end{figure}

The third long series we consider is the mortality rate 
F for atlantic cod. It has a strong and almost linear 
persistence in time, see Figure \ref{figure:F:data}. 
The raw correlation with $\HSI_{\bulk}$ was $-0.19$, which 
is quite high compared with the others series we consider.
It is however currently not clear whether the effect will 
be present after correcting for a linear trend. 
If analysed on its own, subtracting a linear trend makes 
the resulting residual series look like white noise. 
Morevoer, using a standard Dickey--Fuller test to check for 
unit roots, we keep the null hypothesis (with a p-value about 0.42)  
and can not exclude the possibility of a unit root and hence 
that the series is not stationary. Since the series is used 
as a predictor for the HSI index, such a potential lack of 
stationarity is unproblematic. Studying plots like those 
shown in Figure \ref{figure:Kola:WinterVSAnnual} 
indicate again that last year's mortality rate 
is of more importance than that of the current year. 

\subsection*{(iv) Capelin}

As a proxy for the amount of available food we will use the ratio 
of total stock biomass capelin on total biomass spawning stock
estimated from the years 1980 to 2012.
From univariate analyses we learn that a log-transform 
of this food proxy improves the prediction quality 
with respect to the HSI index. This seems to be 
reasonable, since relative changes in the food supply 
is not the same when there is an insufficient amount of food 
as when in abundance. This also results in a raw correlation 
of size 0.56, indicating a strong linear relationship 
and a reasonable predictive ability.

An autoregressive model of order three gives a 
good fit to the capelin series. This suggests using three distinct 
lags as predictors for the HSI. Including more than one lag, 
however, i.e.~the amount of food last year, does not 
improve on the model. This is a reminder that 
preliminary results associated with univariate analyses 
may be misleading when building joint multivariate models;
see the discussion below.  

\subsection*{(v) Annual sunspot numbers}

The multitalented national hero Fridtjof Nansen 
(polar explorer, athlete, writer, artist, diplomat, 
winner of the Nobel peace prize) was also a broadly 
oriented scientist, publishing in zoology
and oceanography. With physical oceanographer 
Bj{\o}rn Helland-Hansen he published the ambitious 
\textit{The Norwegian Sea: Its Physiological Oceanography, 
Based Upon the Norwegian Researches 1900--1904}
\citep{HellandHansenNansen1909}, which also recorded various 
time series pertaining to deep-sea temperatures, salinity, 
density, seasonal variations, the Polar currents, etc. 
The authors also studied how oceanographic and other parameters 
could be related to the growth and spawning of food-fishes, 
and indeed also examined the liver index. They also appear 
to argue that such measurements are related to and perhaps 
causally influenced by ``the periods of Sun-spots'' (Ch.~VII.3). 
Such ideas of the annual sunspots exerting influence 
on the earth's climate and biology 
had also been examined and speculated over by other scholars, 
such as \cite{Ljungman1879}; see the account 
of \cite[Ch.~5]{Smith94} reviewing this historic period,
and also \cite{Lindquist02, Yndestad09}.

\citet[p.~186]{JHjort14} was however sceptical towards
such viewpoints. He got hold of the sunspot numbers 
for the period 1880--1911 in question, from Otto Pettersson,
displaying them along with the liver index series 
in his Figure 116, and commented,
``they do not, however, by any means coincide'', going
on to state his disagreement: 
``The only warrantable conclusion would seem to be,
that no relation can be shown to exist between the two
phenomena, in any case not for the present, nor in the
way suggested by Helland-Hansen and Nansen.'' 
Incidentally, \citet[Part 4]{Joelle11} in his illuminating
biography of Nansen records and comments on several 
clashes between Nansen and Hjort, regarding matters 
of both research administration, how to conduct science,
and how to communicate research questions and findings 
to the general audience. 

\begin{figure}[ht]
\centering
\includegraphics[width=0.45\textwidth]{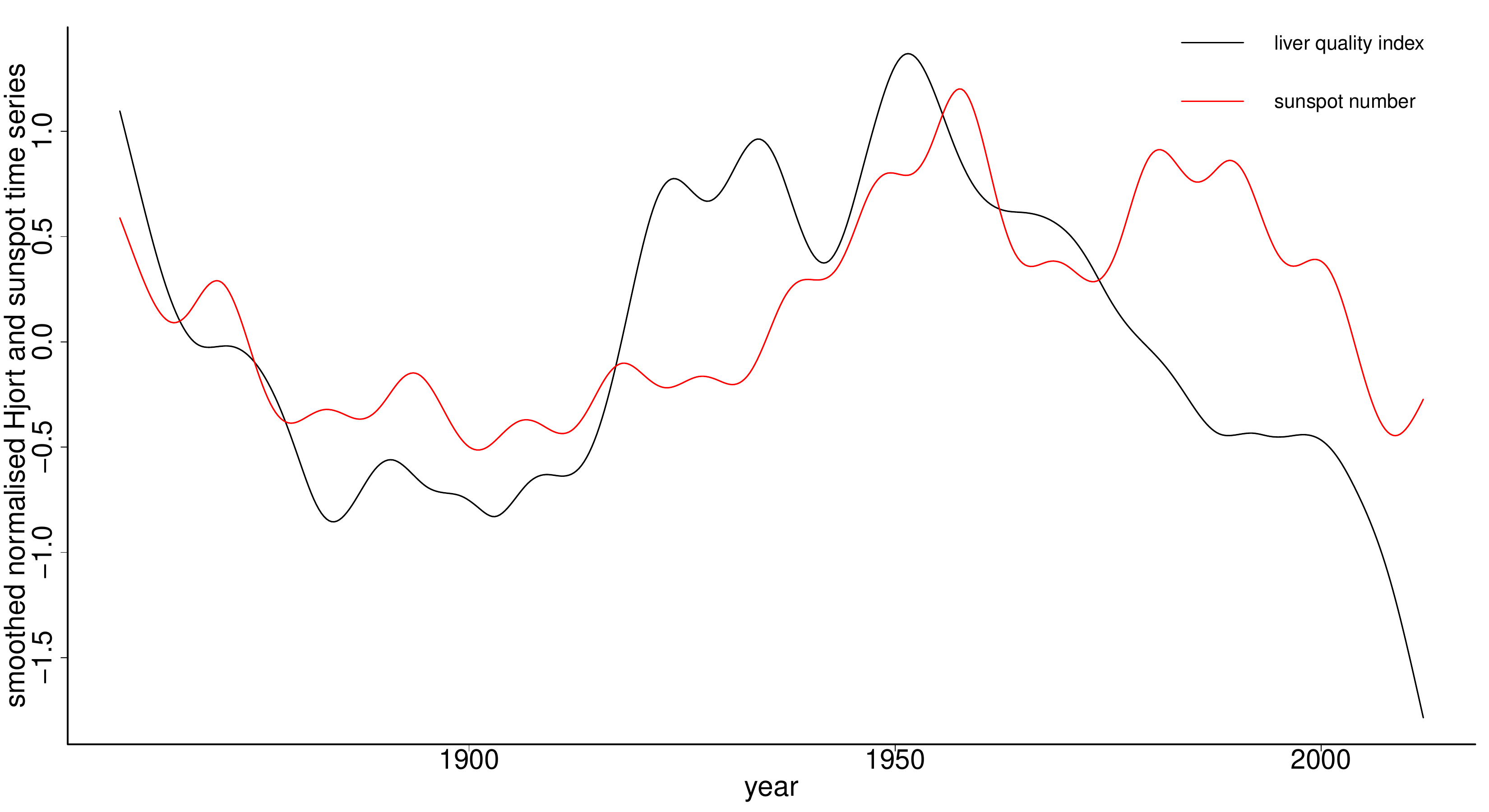}
\caption{The Hjort $\HSI_\bulk$ time series (black) 
with annual sunspot numbers (red) for 1880 to 2012, 
both standardised to have mean zero and unit standard deviation. 
The estimated correlation is $0.11$. This and 
related tests and graphs show little or no signs of any 
underlying relationships.}
\label{figure:HjortAndSunspot}
\end{figure}

A hundred years later we may complement the sunspot 
analysis above, examining the two time series in question for 1859 to 2012
(Figure \ref{figure:HjortAndSunspot}). We find no 
clear statistical relation and side with Hjort over Nansen. 


\subsection*{Multivariate models and multiple covariates}


Let $t_i$ denote year $i$, taken here as calendar year minus 1980, 
and let further $x_{1,i}, x_{2,i}, x_{3,i}$ and $x_{4,i}$ 
be year $i$ observation for the four series discussed above, 
i.e.~average winter Kola temperature, total length, 
mortality rate, and the capelin index 
as a proxy for the food availability. With the univariate analyses
above in mind a natural candidate for a joint model for the liver 
quality index $z_{i}$ is now 
\beq
\label{eq:joint_model}
\begin{split}
z_{i} = \beta_{0} 
+ \beta_{\textrm{year}}  t_i &
+ \beta_{\textrm{kola}} x_{1,i-1}
+ \beta_{\textrm{length}}  x_{2,i} \\
& + \beta_{\textrm{mortality}} x_{3,i-1}
+ \beta_{\textrm{capelin}} x_{4,i-1}
  + \eps_{i}, 
\end{split}
\eeq 
where the $\eps_i$ form a stationary Gaussian 
autoregressive time series of order two. This is a rather 
short series with only 33 complete observations in that 
the capelin series only ranges from 1980 to 2012. 
We therefore supplement model [\ref{eq:joint_model}] 
with a second model, where we bypass the capelin index 
and can use 92 complete observations with $(x_1,x_2,x_3)$ in place.  

As a first analysis we fit the data to 
the regression framework with autoregressive errors of 
order two, to investigate the implied predictive quality, 
see Tables \ref{table:joint_model} and \ref{table:joint_model_capelin}.
A common measure of how well data fit to the model 
is the coefficient of determination, also known as 
$R^{2}$, for the full model above we obtain; 
here $R^{2} = 0.59$ ($R^{2}_{\rm adj} = 0.48$). 
This is in a sense not that representative for the entire series, 
since the capelin series only covers the years 1980--2012. 
For the second analysis which bypasses the capelin index 
we obtain the more promising values of $R = 0.74$ 
($R^{2}_{\rm adj} = 0.72$). 
The baseline autoregressive model (with intercept and linear 
trend) achieves $R^{2} = 0.59$ and ($R^{2}_{\rm adj} =0.58$). 
This indicates that including covariates is 
a real improvement in model quality.  


\begin{table}[ht]
\centering
\begin{tabular}{c|rrl}
& estimate & sd & p-value \\
\hline
$\beta_{0}$   & 2.24 &  2.96 & \;\:\; 0.45 \\
$\beta_{\textrm{year}}$ & -0.04  &  0.02  & \;\:\;  0.06$^{\ast}$ \\
$\beta_{\textrm{Kola}}$  & 0.35 &   0.31 &   \;\:\; 0.26 \\ 
$\beta_{\textrm{length}}$   & 0.03 &   0.03 &   \;\:\;  0.34 \\ 
$\beta_{\textrm{mortality}}$ &  1.14 &    0.67   & \;\:\;  0.11 \\
$\beta_{\textrm{capelin}}$  & 0.25 &    0.10 &  \;\:\; 0.02$^{\ast \ast}$  \\
$\rho_{1}$ & 0.26 &  0.18 &   \;\:\; 0.16 \\
$\rho_{2}$ & -0.54 &   0.18 &   \;\:\;  0.01$^{\ast \ast}$ \\
\end{tabular}
\caption{
Estimates, standard deviation and p-values for the parameters in 
[\ref{eq:joint_model}] fitted using conditional maximum likelihood
estimation, with data from 1980 to 2012.}
\label{table:joint_model}
\end{table}

\begin{table}[ht]
\centering
\begin{tabular}{c|rrl}
& estimate & sd & p-value \\
\hline
$\beta_{0}$   & 1.66 &  1.51 & \;\:\; 0.27\\
$\beta_{\textrm{year}}$ & -0.02  &  0.01  & \;\:\; 0.00$^{\ast \ast}$ \\
$\beta_{\textrm{Kola}}$  & 0.14 &  0.17 & \;\:\; 0.41 \\ 
$\beta_{\textrm{length}}$   & 0.01 &   0.02 &   \;\:\; 0.85 \\ 
$\beta_{\textrm{mortality}}$ &  0.63 &    0.37  & \;\:\; 0.10$^{\ast}$ \\
$\rho_{1}$ & 0.77 &  0.11 &  \;\:\; 0.00$^{\ast \ast}$ \\
$\rho_{2}$ & -0.28 & 0.11 &  \;\:\; 0.02$^{\ast \ast}$\\
\end{tabular}
\caption{
Estimates, standard deviation and p-values for the parameters in 
[\ref{eq:joint_model}] without the capelin index,  
fitted using conditional maximum likelihood estimation,
with data from 1932 to 2012.}
\label{table:joint_model_capelin}
\end{table}

\begin{table}[ht]
\centering
\begin{tabular}{c|rrl}
& estimate & sd & p-value \\
\hline
$\beta_{0}$   & 1.81 &  1.55 & \;\:\; 0.24\\
$\beta_{\textrm{year}}$ & -0.02  &  0.01  & \;\:\; 0.00$^{\ast \ast}$ \\
$\beta_{\textrm{Kola}}$  & 0.14 &  0.17 & \;\:\; 0.41 \\
$\beta_{\textrm{length}}$   & 0.01 &   0.02 &   \;\:\; 0.85 \\
$\beta_{\textrm{mortality}}$ &  0.13 &    0.37  & \;\:\; 0.73 \\
$\rho_{1}$ & 0.68 &  0.11 &  \;\:\; 0.00$^{\ast \ast}$ \\
$\rho_{2}$ & -0.18 & 0.11 &  \;\:\; 0.11$^{\ast}$\\
\end{tabular}
\caption{
Estimates, standard deviation and p-values for the parameters in
[\ref{eq:joint_model}] without the capelin index, 
fitted using conditional maximum likelihood estimation,
with the `corrected' liver quality data series from 1932 to 2012 
from Figure \ref{figure:commercial_survey_combined}.}
\label{table:joint_model_capelin_nils_hsi}
\end{table}

From Table \ref{table:joint_model} we see that in our joint model
the capelin, our proxy for the food availability, is the most 
important covariate. This fits well with our univariate analysis 
above. It is however surprising that the total length index is 
not more vital (this also had a strong correlation with the HSI) 
and seems to be well explained by the other predictors. Note 
that length is even less significant in 
Table \ref{table:joint_model_capelin}. The effect does not seem 
to be a product of the short series (with 33 complete samples) 
used for the analysis in Table \ref{table:joint_model}. 
More important, however, is the fact that $\beta_{\textrm{year}}$ 
is estimated to have a negative slope and with a particularly 
low p-value (in both Tables \ref{table:joint_model} and
\ref{table:joint_model_capelin}). With a somewhat strict 
interpretation this means that the liver quality index is 
currently decreasing with time. This is not an artefact of 
the three successive low $\HSI_\bulk$ values at the end
(2010--2012) of the series (values for which we have already 
expressed mildt scepticism; see the discussion related 
tp Figure \ref{figure:commercial_survey_combined} above); 
indeed we obtain the same result for our robustified series, 
see Table \ref{table:joint_model_capelin_nils_hsi}.

Based on the univariate and joint analyses presented in 
Tables \ref{table:joint_model}, \ref{table:joint_model_capelin},  
and \ref{table:joint_model_capelin_nils_hsi}, we do not believe 
that all covariates, i.e.~year, Kola, length, mortality, and capelin, 
are of the same importance. Moreover, the full model, 
comprising all five predictors and four additional modelling parameters 
($\beta_0$ and three for the autoregressive part) is also 
a bit too much to expect to be able to estimate reliably, 
with only 33 samples.
Traditionally, one would have to do some preselection, 
or model selection, to obtain a subset of predictors to 
use in the final analysis. There are several possibilities, 
with classical approaches including the so-called forward, 
backward or all subsets options. There only significant 
covariates reach the final model, typically evaluated 
via p-values needing to be less than a threshold. 
This is not always the best approach, however, 
since one model is not necessary best for all purposes 
and the relative importance of the different predictors or 
the model complexity may depend on what we are trying to 
answer, e.g.~predictions, threshold probabilities, 
or estimation of underlying structural changes. For 
these reasons the focused model selection strategy (FIC) 
presented above may easily be the best solution.  

\begin{figure}[ht]
\includegraphics[width=0.48\textwidth]{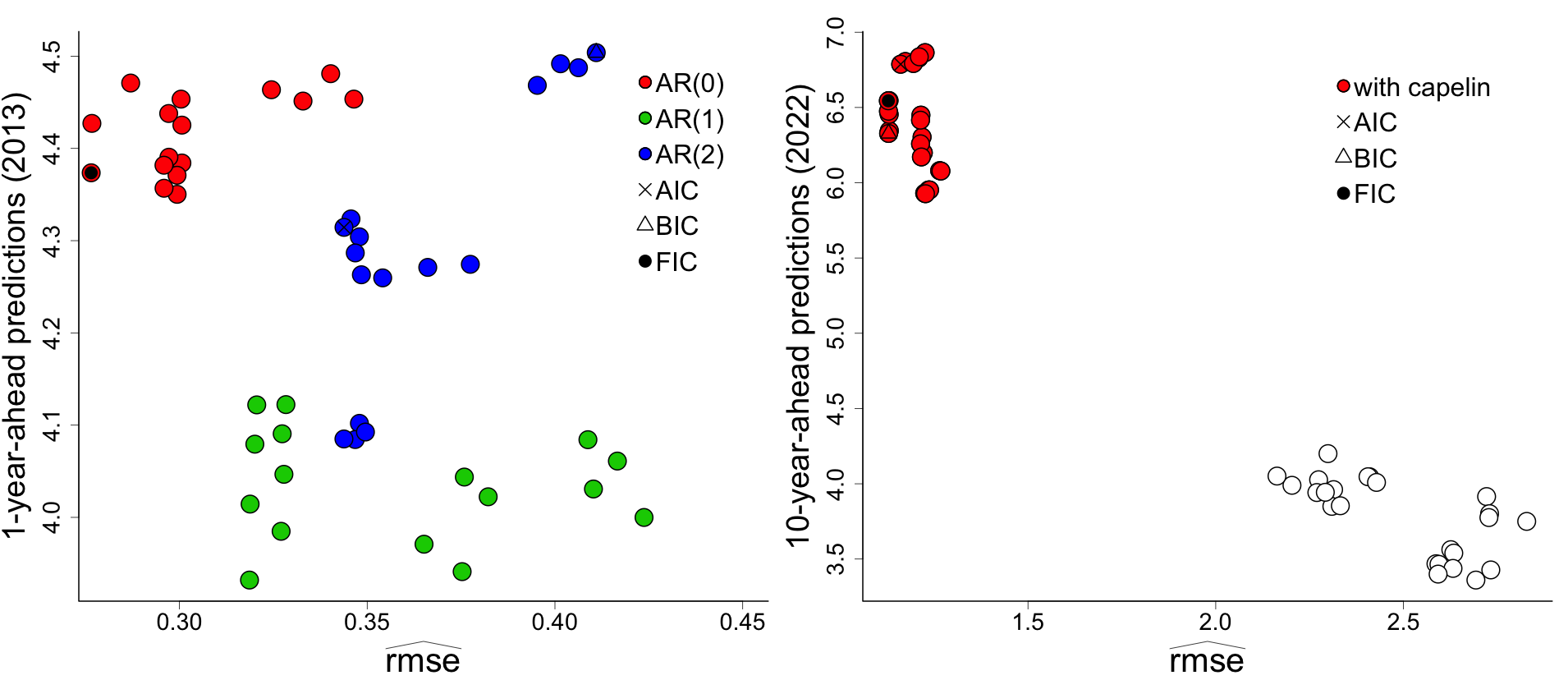}
\caption{
For the same dataset as discussed in Table \ref{table:joint_model}.
For predicting the HSI index for next year (2013) the FIC 
prefers the smallest model with no predictors and independence. 
This seems reasonable, since among the `best' models this is clearly 
the simplest option. It indicates that the FIC, in case of 
a reasonable tie among the candidate model, often selects 
the model with the lowest complexity, which leads to less 
variation in the estimates and also to potentially more precise inference. 
For prediction ten years ahead, the FIC selects the 
independent model using only the capelin series, suggesting 
that the food availability is one of the main signals 
in the long term effect of the HSI index.}
\label{figure:all:FIC:prediction}
\end{figure}

\begin{figure}[ht]
\includegraphics[width=0.48\textwidth]{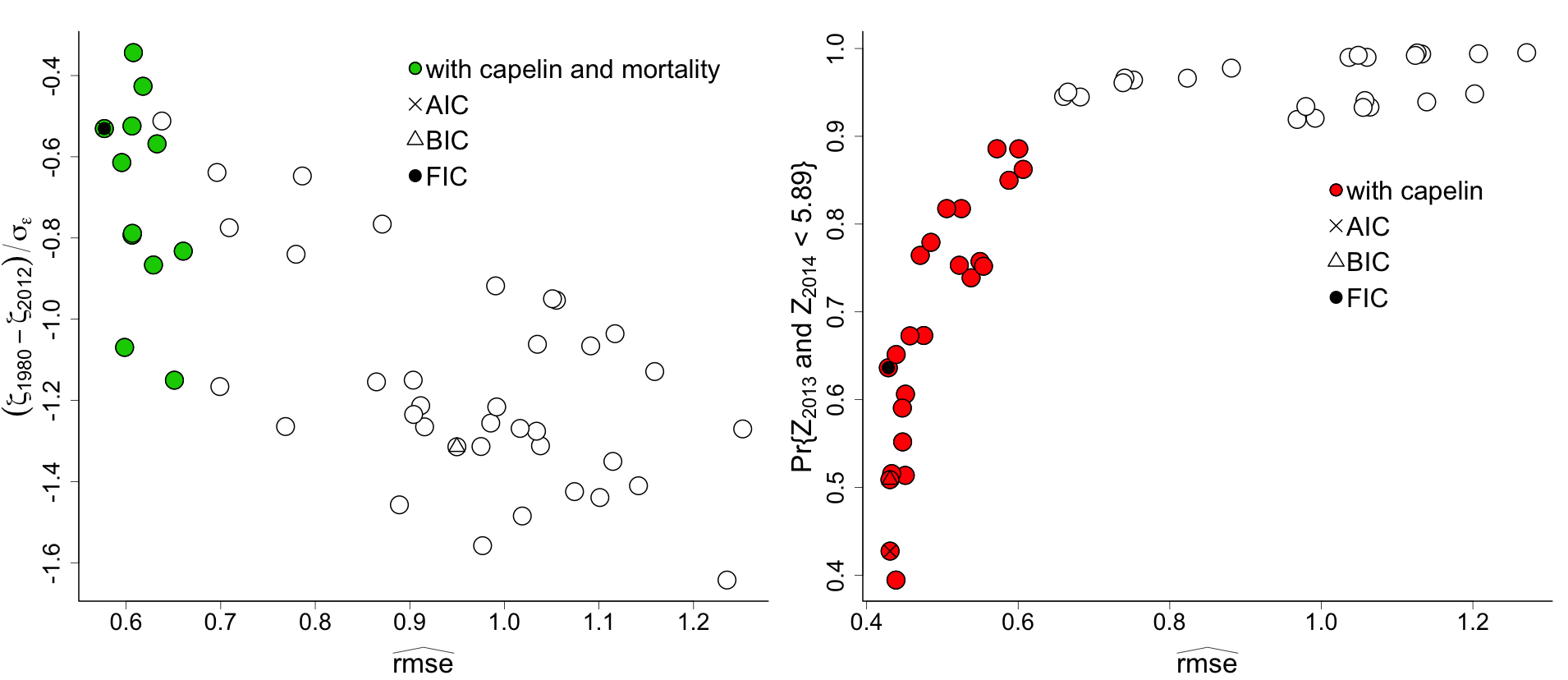}
\caption{
Data are the same as those used to create Table \ref{table:joint_model}.
Slope: Here the FIC agrees with the AIC and prefers the autoregressive model 
of order two with both mortality rate and the capelin index.
Threshold: According to the FIC the best model is the independent 
model that includes all covariates except the Kola temperature.}
\label{figure:all:FIC:more_foci}
\end{figure}

We have already discussed how the FIC can be used to 
find the model best suited for making predictions, 
and above we showed how models of different complexity 
were preferred for different numbers of time steps (years) 
into the future; see Figure \ref{figure:FIC:HSI_with_time}.
There is a variety of further focus functions which 
can help the researcher to select the model best suited 
for the particular problem at hand. 
In Figure \ref{figure:all:FIC:prediction}
we are again attempting to predict the HSI index for 
the future and in Figure \ref{figure:all:FIC:more_foci}
we consider two additional foci:
\beq
\label{eq:three_foci}
\begin{split}
\mu_{\textrm{slope}}& = (\xi_{1980}-\xi_{2000})/\sigma_{\eps}, \\
\mu_{\textrm{threshold}} & = 
   \Pr \{Z_{2013}{\rm\ and\ }Z_{2014}{\rm\ smaller\ than\ } 5.89 \}.
\end{split}
\eeq
Here $\xi_{1980}$ and $\xi_{2000}$ are the expected levels 
of the HSI at years 1980 and 2000, with different formulae
applying for these in different models, and 5.89 is the overall 
mean of the $\HSI_\bulk$ series. 

For all foci the AIC prefers the autoregressive model of order 
two with both mortality rate and the food availability proxy capelin. 
The BIC selects the autoregressive model of order two with 
only the capelin index. 

Note that by using the robustified HSI index series from 
Figure \ref{figure:commercial_survey_combined} (also including 
the capelin), the overall picture is somewhat changed. 
The prediction for 2013 increases to 5.1 (based on FIC, whereas 
the AIC and BIC suggest that it will be as high as 5.3). 
Also, the threshold probability is reduced to 0.22 where the 
model selected by AIC gives an estimate of 0.25 
and BIC suggests a model leading to a estimated probability 
as low as 0.12. 
In general, using the FIC we are able to reach reliable 
conclusions via the most relevant model selection process.

\section*{Combining information across data sources}
\label{section:CDs} 

For various substantive sciences it is important to be able
to pool information sources together, leading to proper
combination or meta-analyses for the more crucial research 
questions. Typical scenarios met e.g.~in the biomedical
sphere involve similarly structured experiments, carried
out by different research groups, after which a meta-analysis
properly combines the summary statistics across these 
individual studies. Increasingly, such meta-questions 
arise also in other fields, and in situations where the 
information sources are rather more diverse. A case
in point is \cite{Myers01}, who analysed a compilation
of over 700 populations of fish, in a framework of
several multivariate time-series. Other and yet more 
complex meta-studies could involve field studies, 
measurements from satellites, mark and recapture data, 
biological evolutionary theories, etc. 
Traditional meta-analysis methods would often not 
cope with the associated data-summary problems, 
also since the individual data summaries themselves would 
differ in format and level of precision.

A powerful general framework for such modern meta-analyses
involves that of {\it confidence distributions}
\citep{XieSingh13, SchwederHjort02, SchwederHjort15}. 
A confidence distribution for a parameter, say $\theta$, 
based on data, say $y$, is a cumulative distribution function 
$C(\theta,y)$ with the property that it spans all confidence intervals;
thus $[C^{-1}(0.05,y),C^{-1}(0.95,y)]$ is a 90\% confidence
interval, etc. This is close in spirit to the Bayesian 
machinery, with a distribution for the parameter of interest, 
but is frequentist, with no subjective priors placed
on the model parameters. See the above references for 
broad discussions and illustrations. It is also useful
to work with the associated {\it confidence curves}
$\cc(\theta,y)=|1-2\,C(\theta,y)|$. These have the 
property that setting $\cc(\theta,y)$ equal to 
a given confidence level, say 0.95, leads to a lower
and an upper solution point, spanning the confidence 
interval in question. Also, the confidence curve 
`points to' the median confidence estimate,
$\hatt\theta=C^{-1}(\half,y)$, which is sometimes but
not always identical to the associated maximum likelihood 
estimate. 

\begin{figure}[ht]
\centering
\includegraphics[width=0.45\textwidth]{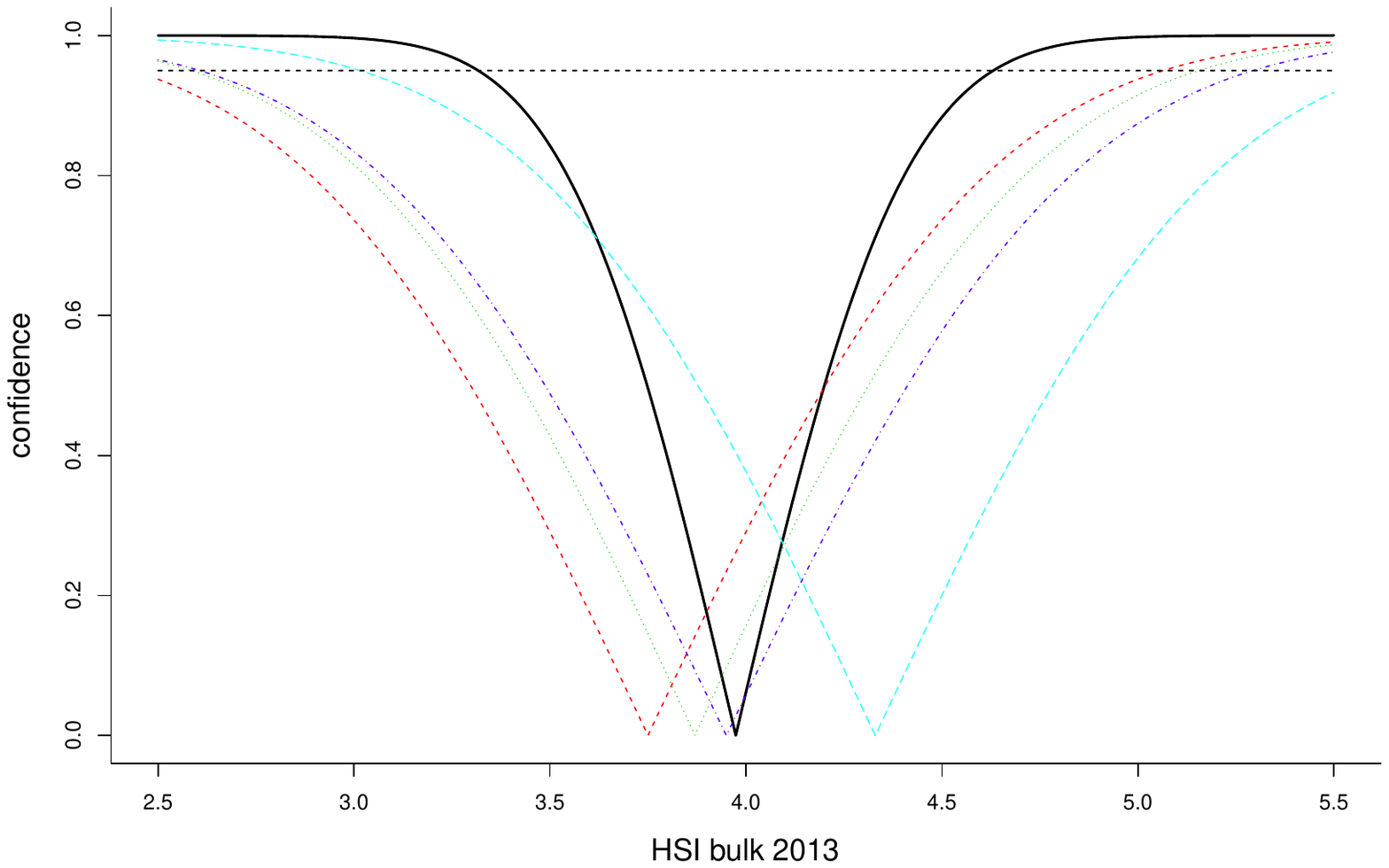}
\caption{
Confidence curves for predicting the value of $\HSI_\bulk$
for the year 2013, based on separate analyses of HSI against
Kola winter temperature, length data, mortality figures,
and Caplin score, along with the optimal confidence curve
combination of these sources. The y-axis indicates degree
of confidence, with the 0.95 confidence line plotted.} 
\label{figure:CDcombination}
\end{figure}

Figure \ref{figure:CDcombination}
displays such confidence curves for predicting 
$\HSI_\bulk$ for the year 2013 (the year after the current
last year of the HSI series), for each of the separate studies
related to Kola winter temperature, mortality rate, 
Capelin score, and length data. 

The general meta-analysis idea associated with confidence 
distributions is as follows. For a parameter of primary
interest, like the trend parameter $\beta_1$ 
of [\ref{eq:hsi_prototype_model}] reflecting a potential
change of the HSI over time, each separate source of 
relevant information leads to a confidence distribution
for the parameter, say $C_1(\beta_1),\ldots,C_k(\beta_1)$. 
Each of these studies might be a complex statistical affair,
regarding model building, interpretation, data gathering, 
and operational practicalities, and might also involve 
various other statistical parameters along the way. The 
confidence distributions may be converted to profiled 
log-likelihood functions, say $\ell_1(\beta_1),\ldots,\ell_k(\beta_1)$,
in ways developed and illustrated in \citet{SchwederHjort15}. 
This leads to the intended meta-analysis by adding up
the log-likelihood pieces and reverting the result to 
a confidence distribution again.

We give two illustrations here. The first has actually 
already been pointed to, when we in 
Figure \ref{figure:commercial_survey_combined} provided 
improved estimates of the HSI bulk index for 1997 to 2012,
combining the observed $\HSI_\bulk$ numbers with 
appropriately transformed versions of the $\HSI_\ind$ numbers 
from the Lofoten survey. The resulting `combined information curve' 
(the black line in the figure mentioned) has emerged via 
the above recipe, involving assessment of each contributing
curve's precision level. Our second illustration 
is Figure \ref{figure:CDcombination}. Here the black curve 
gives the proper optimal combination of the four
other confidence curves, those based on separate analyses
of Kola winter temperature, length data, mortality,
and Caplin score. 95\% confidence intervals may be 
read off, for the separate analyses as well as for the 
combined one. These precision intervals are 
$[2.44,5.06]$ for the Kola winter temperatures, 
$[2.58,5.16]$ for the length data,
$[2.62,5.29]$ for the mortality,
$[3.02,5.64]$ for the Capelin index,
and finally the much shorter interval $[3.32,\allowbreak4.63]$
for the method that optimally combines these 
pieces of information.  


\section*{Concluding remarks}
\label{section:concludingremarks} 

We end our paper with a list of concluding comments, some
pointing to the fruitfulness of further investigations,
regarding both methodology and the examination of 
the Hjort index and its related factors. 

\medskip
\textit{A. To explain or to predict?}
\cite{Breiman01} makes the case that there is a Snow-like
`two cultures' aspect of modern statistical science, 
with one camp eager to fit data in order to make good
classifications and predictions, the other focusing more
on the finer details of models in order to calibrate 
understanding of say underlying biological mechanisms. 
The `to explain or to predict' question, 
cf.~\cite{Shmueli10}, is also pertinent when it comes to 
the many aspects of understanding the Hjort liver quality index.
It is certainly of intrinsic value to understand 
a determining aspect of how the Kola winter temperature
influences the liver quality, even though this might
not shed light on its own on how the size of the 
population develops over the next twenty years.  
As argued in \cite{ClaeskensHjort03, ClaeskensHjort08}, 
it is not a paradox that one model does a good job with 
`explanation' with another one is better at `looking ahead'.
The FIC methods for model selection aids both aims. 

\medskip
\textit{B. Influential factors, changes, fluctuations: deeper issues.}
Regarding the `to explain or to predict' dimensions pointed 
to in the previous remark, we suggest that Johan Hjort 
expressed deep interest in both. His work reflects energetic 
fascination with both the finer aspects and details of how 
nature works and how insights lead to assessment of 
the future, from life in the ocean to human endeavours 
and industries associated with it. He was also profoundly
interested in the nature and business of variation 
and fluctuations, in and by themselves, also in his 
more philosophical writings (e.g.~\citet[Ch.~III]{JHjort20}),
seeing also tentative consequences for the sociology
of human beings.   


He was also clearly interested in fluctuations, and in 
their interplay and correlations with other phenomena,
long before his 1914 book. In the other classic 
\textit{The Depths of the Ocean}, with Sir John Murray
\citep{JHjort12}, he studied percentage of fat in sprats 
caught off the Norwegian west coast in different months, 
comparing these with average temperature of the surface 
of the sea, off Bergen, in each month of the year, 
and comments: 
``The fat-contents of the sprat increase during summer,
when there is a rise in temperature, while both decrease
towards the end of the year; it follows from this that
the growth of the fish must be influenced by the prevailing
temperatures in different waters.'' The remark captures
the spirit of constructive curiosity that we still need 
in order to learn more. 


The following quotation is also apt, for our limited
efforts and successes in understanding how other factors
influence the lives, quantity and quality of 
the skrei (from \citet[Ch.~VI]{JHjort38}, the chapter 
on the methods of correlation and experiment): 
``The cause of our anxiety is the critical fear of an 
overestimation of the method, or rather the conclusions 
drawn from its application. The correlations which we 
call adaptation are certainly of far too great a 
complexity for any postulate of causal connections, 
for the organism and environment are two enormously 
complicated conceptions. The understanding of the 
causal connections between them demands first of all 
a prolonged critical analysis like that which Gegor 
Mendel introduced into the biological study of the 
external characters of the plant, but an analysis 
of this kind is possible only on the basis of previous 
experience created by studies of correlations.'' 



\medskip
\textit{C. Model averaging.}
Sometimes it is fruitful to not merely point to one
winning statistical model (and to discard all its competitors),
but to keep several good models on board. Model averaging
is the term for keeping several model based estimates,
with the final analysis being a mixture across these.
The weights in question, given to the different models,
might be based on FIC scores, so that the best looking models
have higher weights than those not scoring well for the 
purpose at hand. This is sometimes particularly fruitful 
in prediction settings
\citep{HjortClaeskens03, ClaeskensHjort08, Hansen08, ChengHansen14}.

\medskip
\textit{D. Time-varying autoregressive modelling 
via locally stationary processes.} 
The time series models we have used in this paper 
are of the type `trend function linear in covariates 
plus low-order autocorrelated noise'. These are effective,
not difficult to work with, and reasonably robust, in that
moderate deviations from the model used will not severely 
disturb inferences. It is neverthless fruitful to attempt
other variations and perhaps more sophisticated time series
models, e.g.~involving a standard deviation function 
$\sigma_t$ varying with year $t$. Similarly, some 
of the bridge monitoring model checks we have carried
out for the HSI and related data series indicate that 
model parameters may change over time. A pertinent
class of models able to cope with such aspects is that 
worked with in \cite{Dahlhaus97}, the time-varying 
autoregressive model (tvAR).
These are defined mathematically by  
\beqn
Y_{n, i} + \sum_{j=1 }^{p} \alpha_{j}(t/n) Y_{n, i - j} 
   = \sigma(t/n) \eps_{t}, \quad t \in \mathbb{Z}, 
\eeqn
where $\eps_t$ are independent and standard normal. When 
the $\sigma(\cdot)$ and $\alpha_j(\cdot)$ functions are
constant, we are back to the familiar ground of 
autoregressive time series. We have used these models 
for the HSI bulk series, using order $p=2$ above, 
and found that the $\sigma(\cdot)$ function is not 
constant over time, whereas the $\alpha_1(\cdot)$ 
and $\alpha_2(\cdot)$ coefficient functions are approximately
constant. 



\section*{Acknowledgements}

We are grateful to Jennifer Devine and Jon Egil Skj{\ae}raasen
at the Hjort Centre for Marine Ecosystems Dynamics 
and the Institute of Marine Research (IMR, Bergen), 
and also to Bjarte Bogstad at the IMR, 
for making relevant datasets available to us 
as well as for fruitful discussions. 
The Kola temperature time series stems from systematic work 
at the Polar Research Institute of Marine Fisheries and Oceanography 
(PINRO, Murmansk) reported on in \cite{Boitsovetal12}, with the 
data kindly provided by Randi Ingvaldsen at the IMR 
who communicates with PINRO in these regards. 
G.H.H.~and N.L.H.\allowbreak~are also indebted to the Research Council 
of Norway for partial funding of the five-year project 
FocuStat (Focus Driven Statistical Inference With Complex Data),
led by Hjort. 

\bibliographystyle{cjfas}
\bibliography{johanhjort_bibliography14}

\end{document}